\documentclass[aps,prl,superscriptaddress,twocolumn,amsmath]{revtex4-2}

\usepackage[utf8]{inputenc}
\usepackage{graphicx}
\usepackage{amsfonts,amssymb,amsmath}
\usepackage{siunitx}
\usepackage{mhchem}
\usepackage{xcolor}

\begin{document}

\title{Current-induced breakdown of the quantum anomalous Hall effect}

\author{Gertjan Lippertz}
\thanks{These authors contributed equally to this work}
\affiliation{Physics Institute II, University of Cologne, Z\"ulpicher Str. 77, 50937 K\"oln, Germany}
\affiliation{KU Leuven, Quantum Solid State Physics, Celestijnenlaan 200D, 3001 Leuven, Belgium}

\author{Andrea Bliesener}
\thanks{These authors contributed equally to this work}
\affiliation{Physics Institute II, University of Cologne, Z\"ulpicher Str. 77, 50937 K\"oln, Germany}

\author{Anjana Uday}
\thanks{These authors contributed equally to this work}
\affiliation{Physics Institute II, University of Cologne, Z\"ulpicher Str. 77, 50937 K\"oln, Germany}

\author{Lino M.C. Pereira}
\affiliation{KU Leuven, Quantum Solid State Physics, Celestijnenlaan 200D, 3001 Leuven, Belgium}

\author{A. A. Taskin}
\email[]{taskin@ph2.uni-koeln.de}
\affiliation{Physics Institute II, University of Cologne, Z\"ulpicher Str. 77, 50937 K\"oln, Germany}

\author{Yoichi Ando}
\email[]{ando@ph2.uni-koeln.de}
\affiliation{Physics Institute II, University of Cologne, Z\"ulpicher Str. 77, 50937 K\"oln, Germany}

\begin{abstract}
The quantum anomalous Hall effect (QAHE) realizes dissipationless longitudinal resistivity and quantized Hall resistance without the need of an external magnetic field. However, when reducing the device dimensions or increasing the current density, an abrupt breakdown of the dissipationless state occurs with a relatively small critical current, limiting the applications of the QAHE. We investigate the mechanism of this breakdown by studying multi-terminal devices and identified that the electric field created between opposing chiral edge states lies at the origin. We propose that electric-field-driven percolation of two-dimensional charge puddles in the gapped surface states of compensated topological-insulator films is the most likely cause of the breakdown.
\end{abstract}
\maketitle

%
%

\begin{figure*}[t]
\centering
\includegraphics[width=17cm]{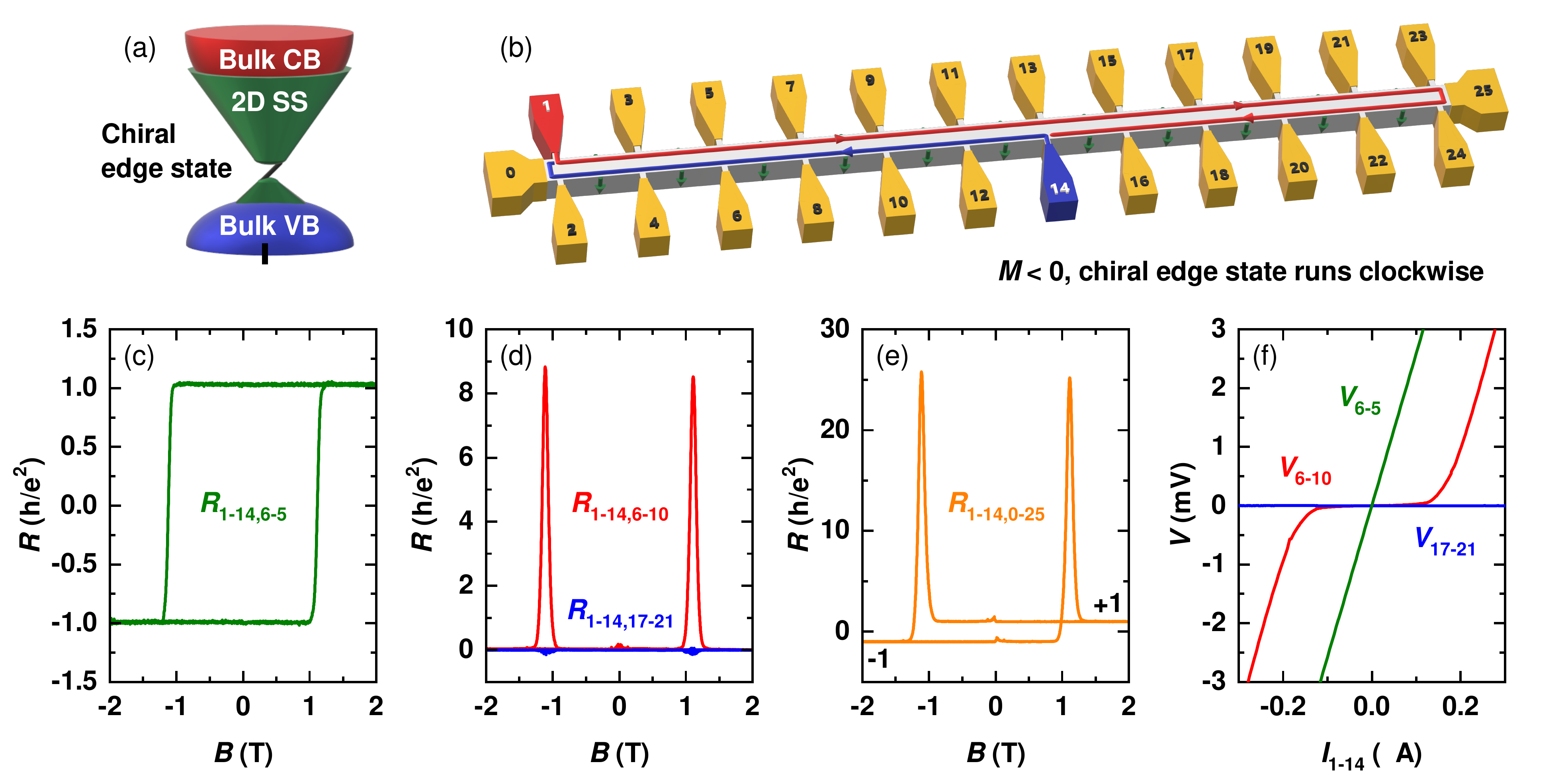}
\caption{Local and nonlocal transport measured below the breakdown current in device A. (a) Illustration of the energy spectrum at the edge of the sample. Bulk conduction band (CB) and valence band (VB) are well separated and the 2D surface-state (SS) spectrum is gapped; when the chemical potential is located inside the gap, the current flows solely through the chiral edge state. (b) Schematic picture of the 26-terminal Hall-bar device with the width of 100~\si{\um}. The red (blue) line shows the high (low) potential portion of the chiral edge state for a downward, out-of-plane magnetization ($M<0$) when current flows from contacts 1 to 14; arrow heads signify the direction of the current flow. (c-e) Magnetic-field dependence of the resistance measured between contacts 6-5 (c), 6-10 and 17-21 (d), and 0-25 (e), showing the QAHE in the local and nonlocal regions at 15~mK with a DC current of 10~nA. (f) Four-terminal current-voltage characteristics in the local and nonlocal regions, measured at 15~mK in +2~T. The breakdown of the QAHE occurs in the local region (voltage between contacts 6-10) at $\sim$0.16~\si{\uA}.
}
\label{fig:nonlocal}
\end{figure*}

The quantum anomalous Hall effect (QAHE) \cite{Yu2010} has been achieved in thin films of the topological insulator (TI) material (Bi$_x$Sb$_{1-x}$)$_2$Te$_3$ doped with Cr or V \cite{Yu2010,Chang2013,Chang2015A}, where spontaneous magnetization $M$ perpendicular to the surface opens an exchange gap in the two-dimensional (2D) surface states [Fig.~\ref{fig:nonlocal}(a)]. The hallmark of the QAHE is the dissipationless longitudinal transport accompanied by quantized Hall resistance of $h/e^2$. Since the chemical potential of the 2D surface states must be tuned into the exchange gap to realize the QAHE, the system is called quantum anomalous Hall insulator (QAHI) in which chiral edge states are the only transport channels. The QAHI is a promising platform for novel quantum phenomena \cite{Breunig2021}, including topological magnetoelectric effects \cite{Qi2008} and chiral topological superconductivity \cite{Qi2010}.

However, the actual nature of the QAHI is still not well understood: For example, the nonlocal transport in the QAHE has shown deviations from an ideal behavior \cite{Kou2014, Chang2015B}, which was explained by employing the coexistence of additional, helical edge states \cite{Wang2013, Kou2014, Chang2015B}. Also, it has been observed that the dissipationless state breaks down with a much smaller critical current than in the integer quantum Hall effect (QHE), and the origin of this premature breakdown is under significant debate \cite{Kawamura2017, Fox2018, Gotz2018, Okazaki2020}. The breakdown current was reported to scale linearly with the sample width \cite{Fox2018}, which is problematic for applications of the QAHE in mesoscopic devices such as those to braid Majorana zero modes generated in the proximity-induced superconducting state of QAHI \cite{Beenakker2019, Adagideli2020, Hassler2020}.

In this work, we investigate the breakdown of the QAHE in both local and nonlocal measurement geometries to gain insights into its mechanism. Our detailed nonlocal transport data allow us not only to dismiss the contribution of additional edge states, but also to identify the transverse electric field $E_{yx}$ as the driving force for the breakdown. To understand the very low breakdown current, we propose that percolation of charge puddles through the 2D bulk causes the abrupt onset of dissipation at the critical value of $E_{yx}$. We demonstrate that the effect of the current-induced breakdown can be minimized by spatially separating the source and drain contacts from the voltage probes, ideally in a nonlocal measurement configuration. We also show that the breakdown can result in a large spurious contribution in the three-terminal configuration with a narrow contact, which should be avoided in designing the experiments.

The experiments were carried out on thin films of \mbox{V-doped} (Bi$_x$Sb$_{1-x}$)$_2$Te$_3$, grown by molecular beam epitaxy (MBE) on InP (111)A. The InP substrate was kept at 190$^\circ$C while V, Bi, Sb, Te were co-evaporated to produce uniform films of $\sim$8~nm thickness. To tune the chemical potential into the magnetic exchange gap, a Bi:Sb beam-equivalent-pressure ratio of 1:4 was used. The films were protected from degradation in air by depositing a 3-nm Al$_2$O$_3$ capping layer using atomic layer deposition immediately after taking the samples out of the MBE chamber. The films were patterned into multi-terminal Hall-bar devices using photolithography and chemical wet etching. Devices A, B, C are long 26-terminal Hall-bar devices, while devices D and E are regular 6-terminal Hall-bars (see Supplementary Material \cite{SM} for details). The metal contacts were fabricated by sputtering 5~nm Pt and 45~nm Au. All the reported films showed a clean QAHE without the need of gating. 

To clearly disentangle dissipative bulk current paths from the edge transport, we investigate 26-terminal Hall-bar devices, as shown in Fig.~\ref{fig:nonlocal}(b). In the measurement shown in Fig.~\ref{fig:nonlocal}(c-e), the current flowed from contact 1 to 14; namely, a voltage is applied to contact 1 and contact 14 is grounded. As a result, the portion of the edge colored red (blue) in Fig.~\ref{fig:nonlocal}(b) is at the source (drain) potential due to the QAHE. From the point of view of local/nonlocal transport, contacts 1 to 14 belong to the `local' transport region, while contacts 15 to 25 are in the `nonlocal' region. The four-terminal resistance $R_{1\text{-}14,6\text{-}10}$ and $R_{1\text{-}14,6\text{-}5}$ correspond to the longitudinal and transverse resistance, respectively; here, the first index denotes the current probes (1-14), while the second index denotes the voltage probes (6-10 or 6-5). The sample shows a clean QAHE with $R_{1\text{-}14,6\text{-}5}$ equal to the von \mbox{Klitzing} constant $h/e^2$ and a vanishing $R_{1\text{-}14,6\text{-}10}$. 

The nonlocal resistance $R_{1\text{-}14,17\text{-}21}$ shows a near perfect zero resistance throughout the magnetic field sweep. This indicates either a near-perfect nonlocal (and dissipationless) edge transport or the absence of current flow in the nonlocal region. If the latter is the case, the potential of contact 25 would always be equal to the drain potential of contact 14, and hence $R_{1\text{-}14,0\text{-}25}$ would be zero for $M<0$. However, the observed $R_{1\text{-}14,0\text{-}25}$ is quantized to $-h/e^2$ for $M<0$, meaning that the potential of contact 25 is equal to the source potential of contact 1. This proves that the nonlocal edge transport is realized.

Having demonstrated a clean QAHE and nonlocal transport, we now address the breakdown of the QAHE with increasing probe current. Figure \ref{fig:nonlocal}(f) shows the voltages appearing at three different contact pairs as functions of the DC probe current $I_{1\text{-}14}$. The longitudinal voltage $V_{6\text{-}10}$ shows a broad plateau at 0~V up to $\sim$0.16~\si{\uA}, above which a sharp increase signifies the breakdown of the dissipationless state. Note that this breakdown current value is among the highest reported so far \cite{Kawamura2017, Fox2018, Gotz2018, Okazaki2020}. The Hall voltage $V_{6\text{-}5}$ follows the expected linear behavior of the QAHE, $V_{6\text{-}5} \approx (h/e^2)I$, with only a little deviation at high current. Obviously, the breakdown in $V_{6\text{-}5}$ is much less pronounced than in $V_{6\text{-}10}$. One can readily show, using the Landauer-B{\"u}ttiker formalism \cite{Buttiker1988}, that a small leakage current crossing the width of the Hall-bar directly affects the longitudinal resistance, while leaving the Hall resistance unaffected (see \cite{SM}). This is consistent with the earliest studies of the QAHE, where the transverse resistance was close to $h/e^2$ while having a sizable longitudinal resistance of several k$\Omega$ \cite{Chang2013}. 

It is important to notice that while the breakdown is clearly observed in the local transport region, it is absent in the nonlocal region: As one can see in Fig.~\ref{fig:nonlocal}(f), $V_{17\text{-}21}$ remains zero beyond $\sim$0.16~\si{\uA}. This points to the transverse electric field as the driving force of the breakdown, as was also suggested in Refs. \cite{Kawamura2017, Fox2018}, because the transverse electric field is absent in the nonlocal region where the edge potential is constant. Moreover, we observed no sign of dissipation in the nonlocal regions (see \cite{SM} for additional data), which speaks against the presence of additional dissipative edge states coexisting with the chiral edge state as was proposed in Refs. \cite{Wang2013, Kou2014, Chang2015B}.

\begin{figure}[t]
\centering
\includegraphics[width=8.6cm]{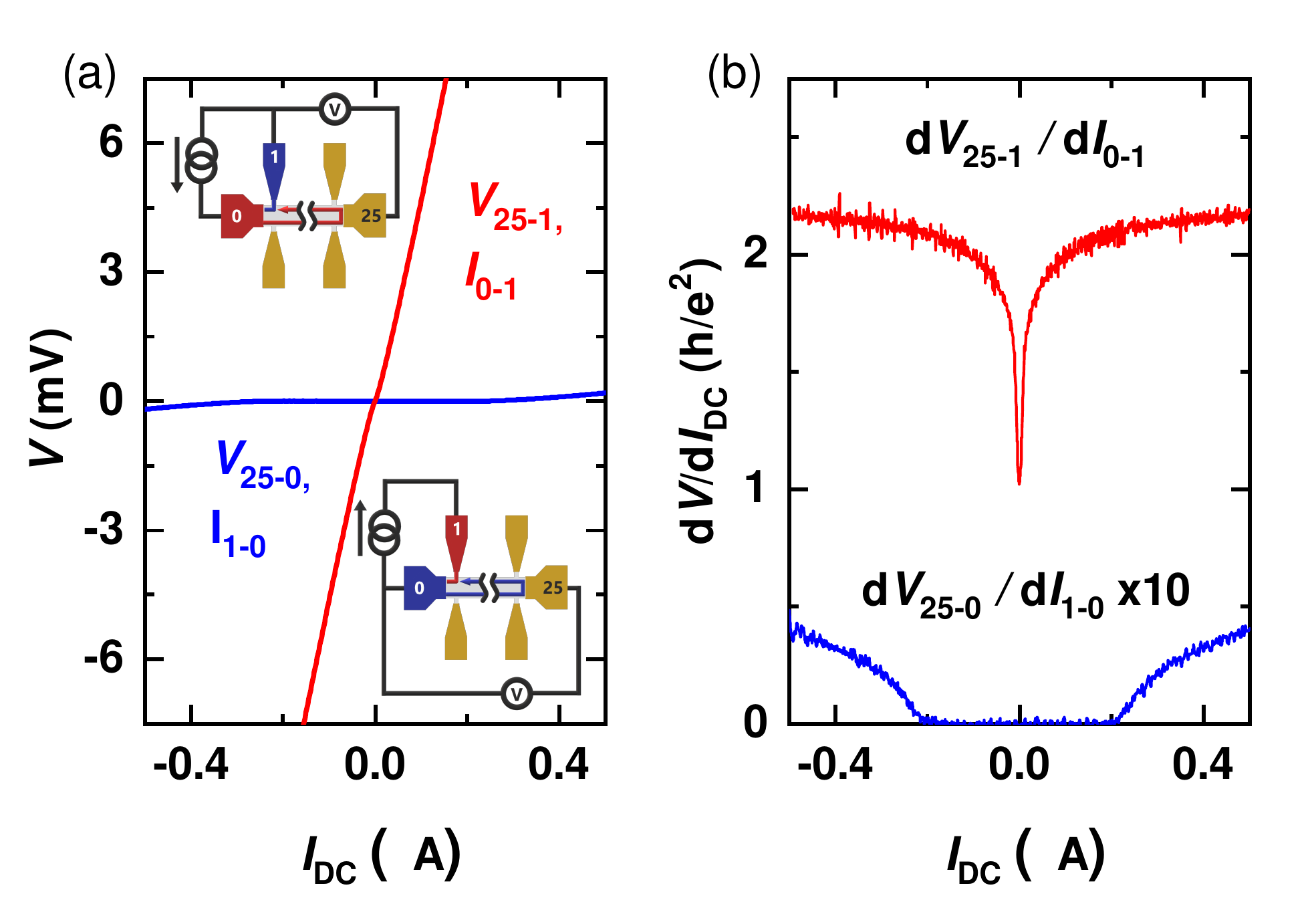}
\caption{The current-induced breakdown in device B for a wide and narrow electrode. (a) The three-terminal current-voltage characteristic of contact 0 (100~\si{\um}, blue) and contact 1 (20~\si{\um}, red) with contact 25 as reference, measured at 40~mK in +2~T ($M > 0$). Insets show the measurement configurations. (b) Corresponding differential resistance for contacts 0 and 1. The curve for $dV_{25\text{-}0}/dI_{1\text{-}0}$ is magnified by a factor of 10 for clarity.
}
\label{fig:contact}
\end{figure}

Since the separation between the high- and low-potential branches of the chiral edge state directly determines the strength of the transverse electric field, one would expect the breakdown to play a strong role near the source and drain contacts where the two branches come together. In our long Hall-bar device shown in Fig.~\ref{fig:nonlocal}(b), contacts 1 to 24 are made via a 20-\si{\um}-wide section of the magnetic TI film (see Fig. S1(b) in \cite{SM} for schematics), while contacts 0 and 25 are made along the full width (100 \si{\um}) of the device. To study the effect of the contact width on the breakdown, we compared the three-terminal resistance involving contacts 0 and 1. Figure \ref{fig:contact} shows the three-terminal $I$-$V$ characteristics and corresponding differential resistance for two configurations with $M > 0$. The voltage $V_{25\text{-}0}$ was measured with contact 0 as the drain ($I_{1\text{-}0}$). Since the edge current flows counterclockwise for $M>0$, contact 25 was at the drain potential in this measurement. Indeed, the differential resistance $dV_{25\text{-}0}/dI_{1\text{-}0}$ is approximately zero with an upturn at $\sim$0.16~$\mu$A due to the breakdown of the QAHE. Hence, there is no additional resistance associated with contact 0.

On the other hand, $V_{25\text{-}1}$ was measured with contact 1 as the drain ($I_{0\text{-}1}$). Now contact 25 is at the source potential, and hence $dV_{25\text{-}1}/dI_{0\text{-}1}$ should be equal to $h/e^2$. However, $dV_{25\text{-}1}/dI_{0\text{-}1}$ immediately deviates from $h/e^2$ with a finite $I_{\text{DC}}$ [see red curve in Fig.\ref{fig:contact}(b)]. As explained in \cite{SM}, this is due to the large electric field appearing in the narrow contact arm, causing an immediate breakdown in the contact arm and enhancing the three-terminal resistance. This demonstrates the necessity to avoid a three-terminal configuration with a narrow contact to minimize the breakdown effect. For example, in a recent study using a $\sim$200-nm-wide Nb electrode on top of a QAHI film, the breakdown of the QAHE was the dominant contribution to the measured conductance \cite{Kayyalha2020}, making it difficult to detect the Andreev reflection.

\begin{figure}[b]
\centering
\includegraphics[width=8.6cm, trim={2cm 8.5cm 2cm 2cm},clip]{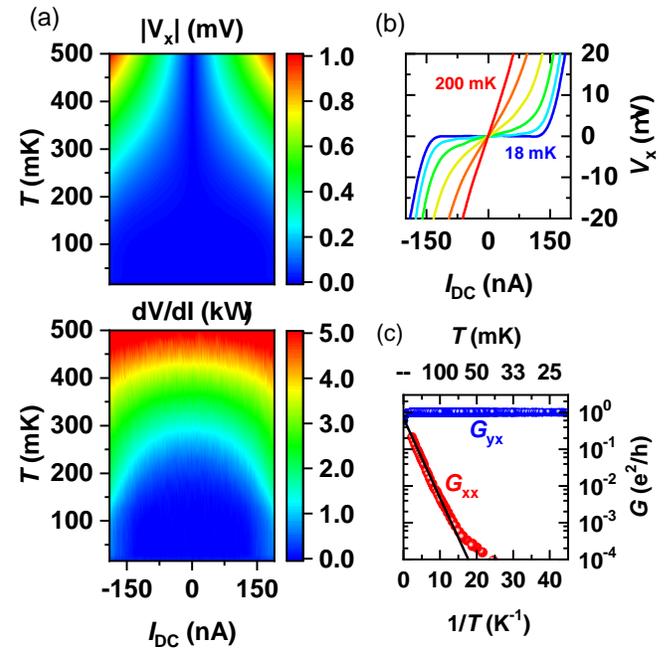}
\caption{Temperature dependence of the QAHE. (a) 2D color mapping of the longitudinal voltage $\lvert V_{x} \rvert$ and the corresponding differential resistance $dV/dI$ as functions of temperature $T$ and current $I_{\text{DC}}$ for device E, measured with 50-$\mu$m-wide contacts at 0~T after training at +2~T . (b) $V_x$ vs $I_{\text{DC}}$ at 18, 100, 125, 150, 175, and 200 mK, showing the evolution of the current-induced breakdown curve. (c) Arrhenius plot of the longitudinal conductance $G_{xx}$ and the transverse conductance $G_{yx}$ of device D, measured with an AC probe current of 10~nA. Black solid line is a fit to the linear behavior $G_{xx} = G_0 e^{-T_0/T}$, yielding $T_0 \approx 0.5$~K. 
}
\label{fig:temp}
\end{figure}

Now we turn to the temperature dependence of the breakdown effect. Figure \ref{fig:temp}(a) shows a 2D mapping of the longitudinal voltage $V_x$ and differential resistance as functions of temperature $T$ and current $I_{\text{DC}}$; the plots of $V_x$ vs $I_{\text{DC}}$ at selected temperatures are shown in Fig.~\ref{fig:temp}(b). A well-extended zero-voltage plateau is seen up to $\sim$100~mK and a linear $I$-$V$ relation is restored at 200 mK. Figure \ref{fig:temp}(c) shows that the thermal activation of charge carriers determines the conductance above $\sim$100 mK with a small activation energy of $k_{\rm B}T_0 \approx 40$~$\mu$eV (i.e. $T_0 \approx$ 0.5~K). This value is comparable to the values $\sim$17--121~$\mu$eV found in previous studies \cite{Bestwick2015, Chang2015B, Kawamura2017,Fox2018,Ou2017,Rosen2017}. It is worthwhile to note that this activation energy is much smaller than the exchange gap of $\sim$14--28~meV observed in scanning tunneling spectroscopy \cite{Lee2015,Chong2020}, hinting at the role of disorder.

\begin{figure}
\centering
\includegraphics[width=8.6cm, trim={1.8cm 12.5cm 3.3cm 2cm},clip]{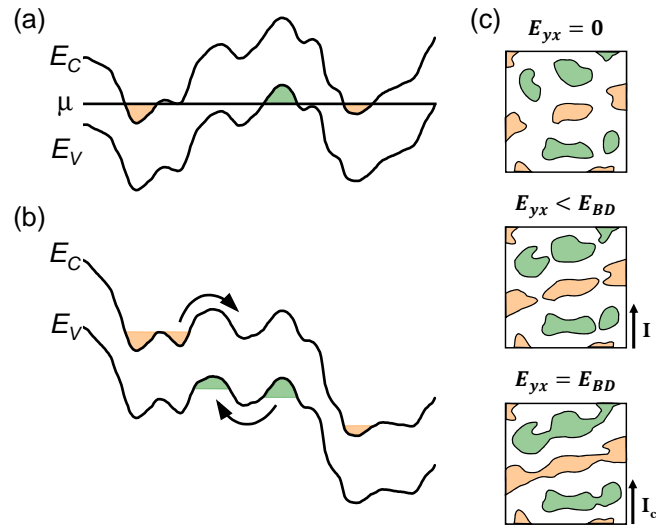}
\caption{Illustration of the puddles in compensated QAHI films and the response to electric fields. (a) Spatial variation of the energy spectrum of the gapped 2D surface state. The meandering lines represent the 2D conduction and valence band edges, $E_C$ and $E_V$, respectively, in the presence of Coulomb disorder. The Fermi level $\mu$ crosses the band edges, creating electron and hole puddles (shaded regions). (b) Spatial variation of the energy spectrum close to the breakdown. Arrows illustrate thermally activated or hopping transport. (c) Growth of puddles driven by increasing electric field $E_{yx}$ until breakdown occurs at $E_{yx}=E_{\text{BD}}$, based on Ref.~\cite{Tsemekhman1997}. The critical current $I_{\rm c}$ is reached with the critical source potential producing $E_{\rm BD}$ between two opposing edge states. The shaded regions correspond to electron and hole puddles in an insulating background.
}
\label{fig:puddle}
\end{figure}

To understand the origin of the breakdown as well as the strongly reduced activation energy, it is useful to consider the role of charge puddles appearing in compensated TI materials \cite{Borgwardt2016, Knispel2017, Breunig2017}. It has been established that puddle formation in three-dimensional (3D) compensated TIs is an unavoidable consequence of the long-range nature of the Coulomb interaction \cite{Skinner2013A, Skinner2012, Chen2013}. While the 3D bulk puddles are strongly suppressed near the surface due to the screening by the metallic surface states \cite{Skinner2013B,Bomerich2017}, 2D surface puddles are predicted to show up in compensated TI thin films \cite{Huang2021}. In this regard, (Bi$_x$Sb$_{1-x}$)$_2$Te$_3$ is a solid-solution of $n$-type Bi$_2$Te$_3$ and $p$-type Sb$_2$Te$_3$, achieving a compensation to result in vanishing 2D surface carriers at low temperature \cite{Zhang2011}. In the QAHI films, the tendency to form 2D puddles would be strong, because the averaged chemical potential is tuned into the gap opened at the charge neutrality point. In such a case, little surface carriers are available to screen the Coulomb potential and the screening can only occur nonlinearly through the formation of 2D electron and hole puddles \cite{Skinner2013A}, as illustrated in Fig.~\ref{fig:puddle}(a). In addition, the large dielectric constant of TI films slows down the decay of the Coulomb potential in space and greatly enhances the puddle formation \cite{Huang2021}. Indeed, signatures of puddle formation have been observed in the resistivity of ultrathin films of (Bi$_x$Sb$_{1-x}$)$_2$Te$_3$ \cite{Nandi2018}. 

In the light of the likely existence of 2D puddles in compensated TI films, we propose that the breakdown occurs via the formation of metallic percolation paths connecting these 2D puddles across the width of the sample. The QAHI films can be thought of as an insulating background containing isolated metallic puddles, as shown in Fig.~\ref{fig:puddle}(c). In analogy to Ref.~\cite{Tsemekhman1997} for the QHE, we propose that at high enough source potential, the insulating regions separating two adjacent electron or hole puddles break down due to the high electric field created between source and drain potentials. Since the local potential is constant within the metallic puddles, the electric field is confined to the insulating regions. As the puddles grow with increasing source potential, the local electric field in-between the puddles increases rapidly, facilitating further puddle growth in a non-linear manner, see Fig.~\ref{fig:puddle}(b-c). At the critical value of the source potential corresponding to the critical current, the growth becomes unstable and leads to an avalanche process \cite{Tsemekhman1997}, so that the metallic paths percolate from one edge of the sample to the other and causes an abrupt onset of dissipation.

The puddle breakdown mechanism proposed here also sheds a new light on the very low temperature required to observe the QAHE, which is much lower than that expected from the Curie temperature $T_{\rm C}$ ($\sim$15--20~K) or the spectroscopically-resolved exchange gap ($\sim$14--28~meV) \cite{Lee2015,Chong2020}. In the presence of charge puddles, electrons are not excited across the 2D exchange gap; rather, electrons and holes are thermally excited from the puddles to the percolation levels \cite{Skinner2013A}. As demonstrated already for 3D bulk puddles in compensated TIs, this reduces the activation energy for thermally-activated transport \cite{Skinner2012, Skinner2013A, Chen2013}. Moreover, at low temperature the electrons and holes may hop or tunnel directly between puddles, possibly giving rise to a crossover from activated transport to variable-range-hopping (VRH) behavior \cite{Skinner2013A}, as was observed in some transport studies on QAHI films \cite{Kawamura2017, Fox2018}.

It is prudent to mention that runaway electron heating \cite{Komiyama2000, Fox2018} and Zenner tunneling \cite{Nandi2018,Huang2021} could provide alternative breakdown mechanisms for the QAHE. However, the heating effect in our QAHI films should be small due to the very low breakdown current values of the QAHE. Also, Zenner tunneling does not result in an abrupt onset of dissipation in the presence of charge puddles, and hence is not consistent with the observation in compensated samples. (More detailed discussions are given in \cite{SM}.)

In conclusion, we demonstrate that the breakdown of the QAHE occurs in the region with the shortest separation between the high- and low-potential branches of the chiral edge state, while it is absent in nonlocal transport regions. This indicates that the transverse electric field is responsible for the breakdown and gives a guiding principle for minimizing the breakdown effect. Moreover, we propose that charge puddles play a key role in the breakdown mechanism for the QAHE and govern the diffusive transport through the 2D bulk states.

\acknowledgements{We thank Y. Tokura, M. Kawasaki, and R. Yoshimi for valuable discussions on the MBE growth of QAHIs. This project has received funding from the European Research Council (ERC) under the European Union’s Horizon 2020 research and innovation program (Grant Agreement No.~741121) and was also funded by the Deutsche Forschungsgemeinschaft (DFG, German Research Foundation) under Germany's Excellence Strategy - Cluster of Excellence Matter and Light for Quantum Computing (ML4Q) EXC 2004/1 - 390534769, as well as under CRC 1238 - 277146847 (Subprojects A04). G.L. acknowledges the support by the KU Leuven BOF and Research Foundation–Flanders (FWO, Belgium), file No.~27531 and No.~52751.}

\clearpage
\onecolumngrid

\renewcommand{\thefigure}{S\arabic{figure}} 
\renewcommand{\thetable}{S\arabic{table}} 

\setcounter{figure}{0}

\begin{flushleft} 
{\Large {\bf Supplementary Information}}
\end{flushleft}

\section{Summary of the devices used in this study}

\begin{figure}[h]
\centering
\includegraphics[width=\linewidth, trim={0.3cm 7.5cm 0.3cm 7.5cm},clip]{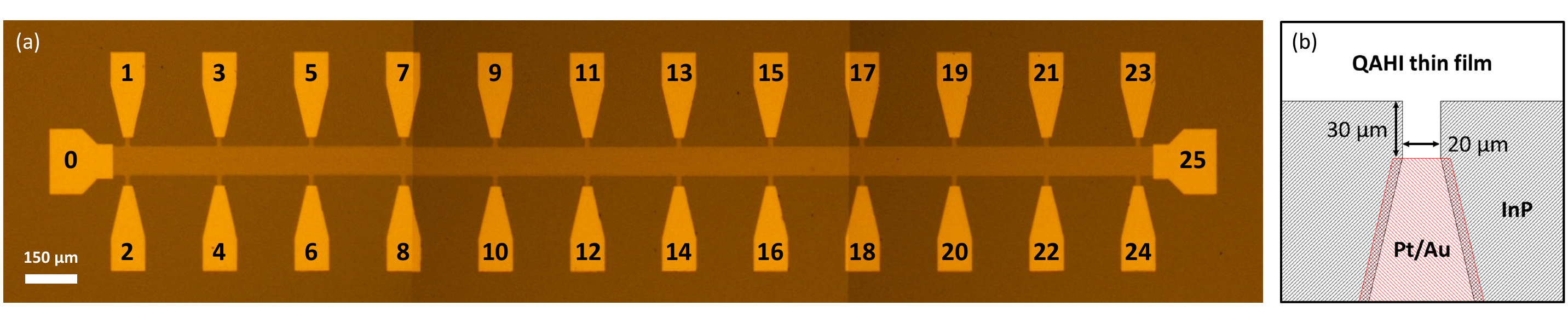}
\caption{(a) Picture of the 26-terminal Hall-bar device~A (width 100~\si{\um}, contact separation 300~\si{\um}). Three photographs are merged, as the device was too long for the microscope camera frame. (b) Schematic of the 20-\si{\um}-wide contacts. 
}
\label{fig:device}
\end{figure}

Five Hall-bar devices were used in this study. Devices A, B, and C are 26-terminal Hall-bar devices, as shown in Fig.~\ref{fig:device}. Devices D and E are regular 6-terminal Hall-bar devices; a picture is included in Fig.~\ref{fig:small4p}(a). All devices showed a clean QAHE. Devices A and B had a breakdown current value of $\sim$160~nA, devices C and D had $\sim$60~nA, and device E had $\sim$120~nA. The width of all the Hall-bar devices was 100~\si{\um}. As can be seen in Fig.~\ref{fig:device}, the contacts from 1 to 24 were made via a relatively narrow, 20-\si{\um}-wide section of the magnetic TI film, whereas contacts 0 and 25 were made along the full width of the device (i.e. 100~\si{\um}). For the 6-terminal Hall-bar devices, the voltage contacts were made via a $50 \times 50$~\si{\um^2} section.

\section{The need for a long multi-terminal Hall-bar device}

Thus far nonlocal transport of the QAHE was only clearly demonstrated by Chang \textit{et al.} in a scratched, mm-size 6-terminal Hall-bar device by flowing a current through neighboring contacts and measuring the three-terminal resistance \cite{Chang2015Bs}. In this section, similar nonlocal measurements are shown for a \si{\um}-size 6-terminal Hall-bar device.

\begin{figure}[h]
\centering
\includegraphics[width=\linewidth]{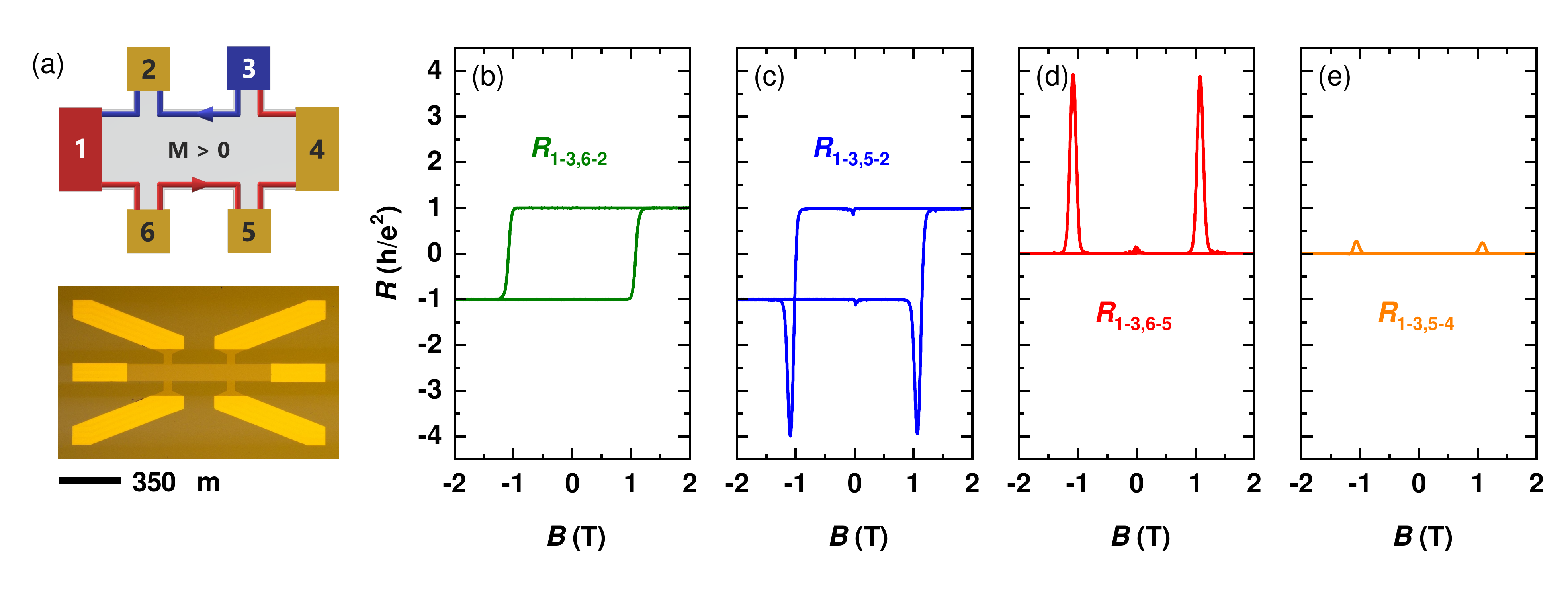}
\caption{(a) Schematic of the high (red) and low (blue) potential portions of the chiral edge state for $M>0$ (top), and a picture of the 6-terminal Hall-bar device D having the width 100~\si{\um} and the contact separation 350~\si{\um} (bottom). (b-d) Magnetic-field dependencies of the `local' resistances $R_{1\text{-}3,6\text{-}2}$, $R_{1\text{-}3,5\text{-}2}$ and $R_{1\text{-}3,6\text{-}5}$, measured at 30~mK with an AC probe current of 10~nA. (e) Magnetic-field dependence of the nonlocal resistance $R_{1\text{-}3,5\text{-}4}$.
}
\label{fig:small4p}\bigskip
\includegraphics[width=\linewidth]{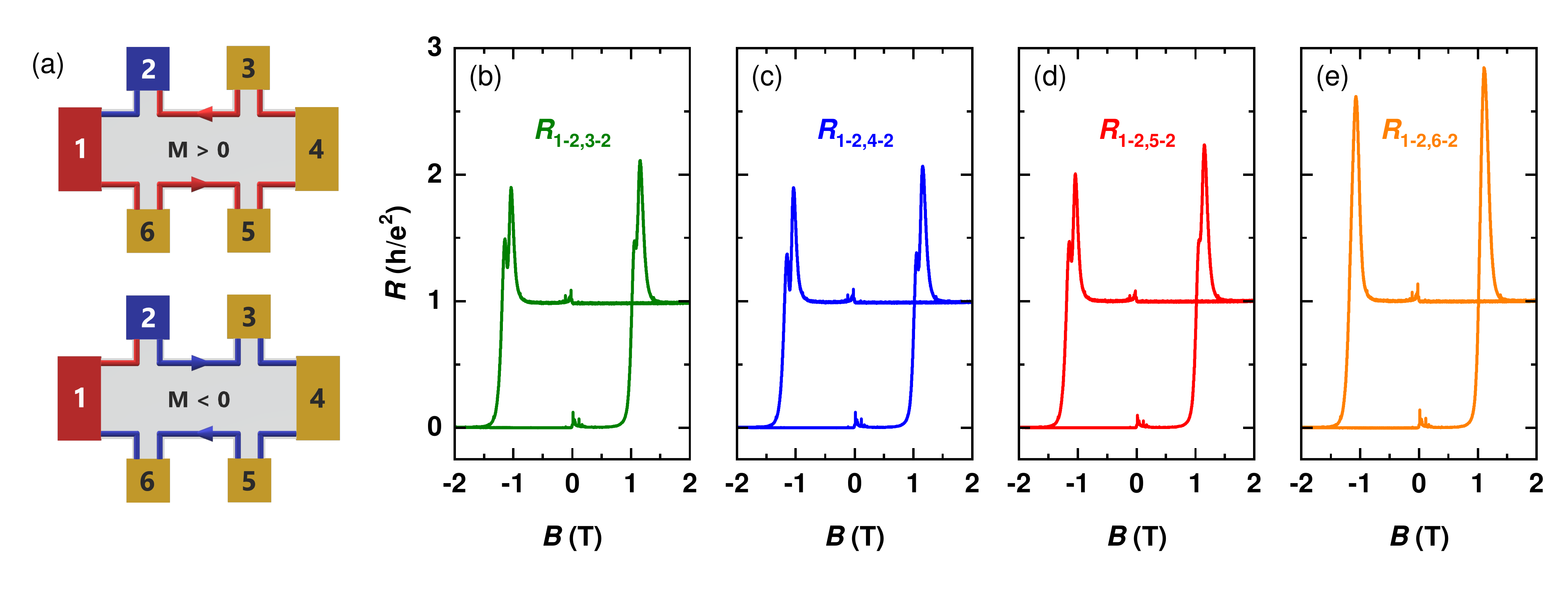}
\caption{(a) Schematics of the high (red) and low (blue) potential portions of the chiral edge state for $M>0$ and $M<0$. (b-e) Magnetic-field dependencies of the resistances $R_{1\text{-}2,x\text{-}2}$ with $x = \{ 3,4,5,6 \}$, measured at 30~mK with an AC probe current of 10~nA.
}
\label{fig:small3p}
\end{figure}

Figure \ref{fig:small4p} shows the four-terminal resistance of our \si{\um}-size 6-terminal Hall-bar device D. The transverse resistance $R_{1\text{-}3,6\text{-}2}$ and the longitudinal resistance $R_{1\text{-}3,6\text{-}5}$ show the quantized hysteresis loop and near-dissipationless state of the QAHE, respectively. Note that $R_{1\text{-}3,5\text{-}2} = R_{1\text{-}3,6\text{-}2} - R_{1\text{-}3,6\text{-}5}$ as expected. The resistance $R_{1\text{-}3,5\text{-}4}$ measures a nonlocal transport and should remain zero for any value of the magnetic field. However, small resistance peaks at the coercive field, corresponding to 2D diffusive transport, are clearly visible. As a result, when the Hall-bar dimension is reduced from mm- to \si{\um}-size, the simple 6-terminal device is no longer suitable for nonlocal transport measurements. For this reason, the Hall-bar in the main text was elongated and extra contacts were added to allow for length-dependent measurements.

Figure \ref{fig:small3p} shows the three-terminal resistance measured in the same nonlocal geometry as in Refs.~\cite{Chang2015Bs, Kou2014s}. The nonlocal resistances $R_{1\text{-}2,x\text{-}2}$ with $x = \{ 3,4,5,6 \}$ show the quantized hysteresis loop superimposed with coercive field resistance peaks. The additional component is due to the small V-doped (Bi$_x$Sb$_{1-x}$)$_2$Te$_3$ sections of the \si{\um}-size source/drain contacts, which add a diffusive contribution to the `nonlocal' three-terminal resistance. Note, however, that for a probe current of 10~nA, no breakdown of the QAHE is observed near the 50~\si{\um}-wide drain contact 2. Detailed analysis of the breakdown of the QAHE are given below, using the Landauer-B{\"u}ttiker formalism.

\section{Toy-model for the current-induced breakdown of the QAHE}

In this section, we describe the effect of the current-induced breakdown of the QAHE on the measured resistance using the Landauer-B{\"u}ttiker formalism \cite{Buttiker1988s}. Our simplistic model by no means describes the physics of the current-induced breakdown fully. However, it does capture the key feature of the breakdown mechanism, namely, the loss of the edge current from the high-potential branch of the chiral edge state to the low-potential branch. Below we show the effect of such a leakage current on the measured four-terminal and three-terminal resistances.

\subsection{Breakdown in the four-terminal measurement geometry}

\begin{figure}[h]
\centering
\includegraphics[width=10cm]{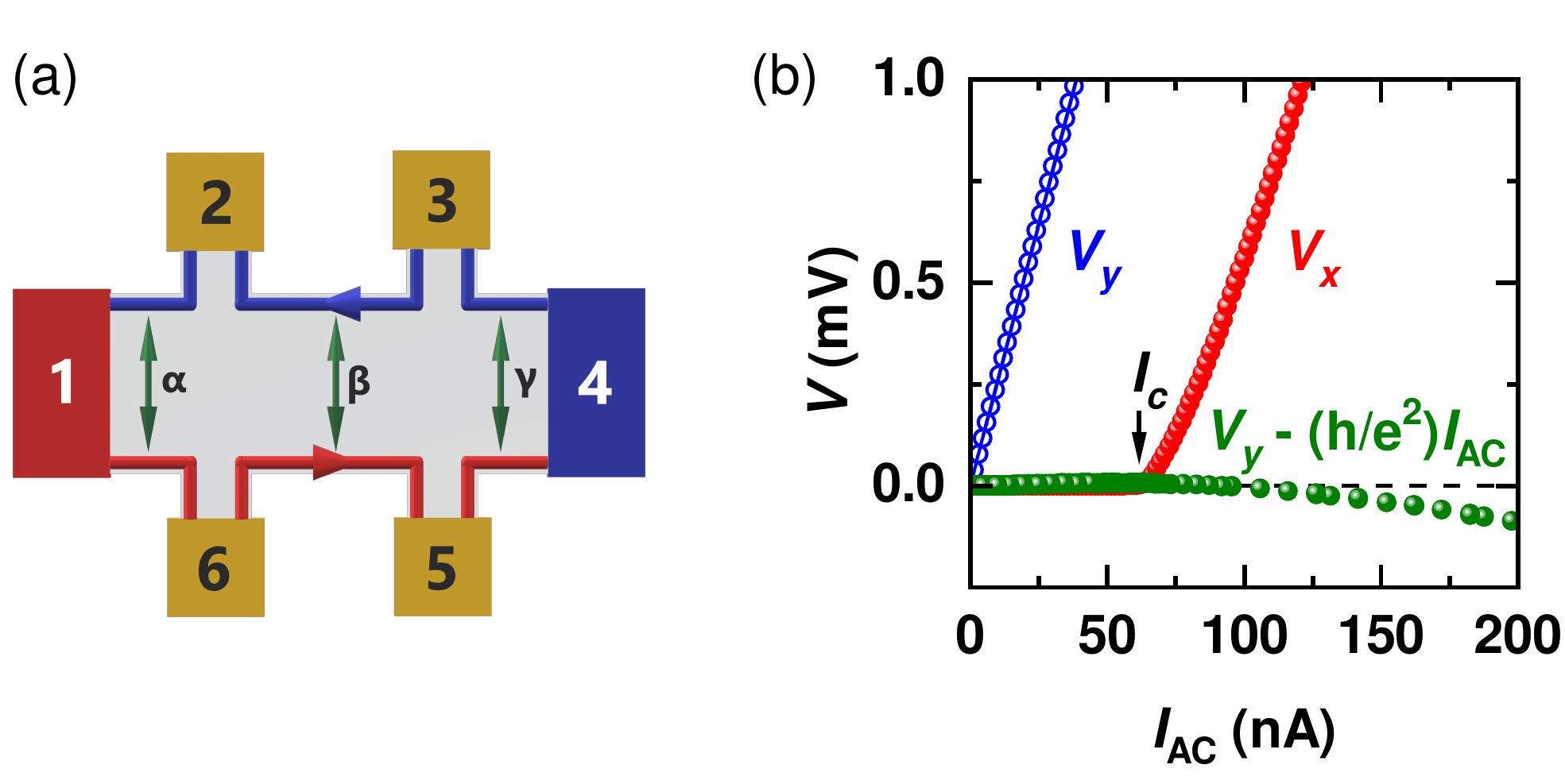}
\caption{Toy-model for the breakdown of the QAHE. (a) Schematic of the edge transport in a 6-terminal Hall-bar device for $M>0$, with the `leakage' between opposing edges parametrized by the scattering probabilities $\alpha$, $\beta$, and $\gamma$. (b) Current-induced breakdown in device~D, measured at 20~mK and 0~T. The sample was trained at +2~T. Above $\sim$60~nA, the longitudinal voltage $V_{x}$ and the transverse voltage $V_{y}$ show deviations from the ideal QAHE.
}
\label{fig:LB4p}
\end{figure}

Figure \ref{fig:LB4p}(a) shows a schematic of a 6-terminal Hall-bar device for an upward, out-of-plane magnetization ($M>0$). The current flows from contacts 1 to 4, and the chiral edge state runs counterclockwise along the sample edge. The high- and low-potential branches of the chiral edge state are separated by the width of the Hall-bar. To describe the fraction of the current leaking from the high- to the low-potential branch, we introduce the scattering probabilities $\alpha$, $\beta$, $\gamma$, as depicted in Fig.~\ref{fig:LB4p}(a). In the Landauer-B{\"u}ttiker formalism \cite{Buttiker1988s}, the current-voltage relation is given by
\begin{equation}
I_i = \frac{e^2}{h}\sum_{j} (T_{ji}V_i-T_{ij}V_j),
\label{eq:LB}
\end{equation}
where $V_i$ is the voltage on the $i$th contact, $I_i$ is the current flowing through the $i$th contact into the sample, and $T_{ji}$ is the transmission probability from the $i$th to the $j$th contact. Consider the edge channel running from contacts 6 to 5; there is a $T_{56} = (1-\beta)$ probability of reaching contact 5 and a $T_{26} = \beta$ probability of scattering to contact 2. Similarly, the other nonzero transmission coefficients can be found to be:
\begin{align}
T_{11}&=T_{62}=\alpha,	&	T_{12}&=T_{61}=(1-\alpha), \nonumber \\
T_{26}&=T_{53}=\beta,	&	T_{23}&=T_{56}=(1-\beta), \\
T_{35}&=T_{44}=\gamma,	&	T_{34}&=T_{45}=(1-\gamma). \nonumber
\end{align}
Using $V_4=0$, $I_1=-I_4=I$ and $I_2=I_3=I_5=I_6=0$, Eq.~\ref{eq:LB} gives a set of equations which can be solved for $I$ and $V_i$. The four-terminal longitudinal and transverse resistances then become:
\begin{align}
R_{xx}&=\frac{V_2-V_3}{I}=\frac{V_6-V_5}{I}=\frac{\beta}{1-\beta}\frac{h}{e^2}, \\
R_{yx}&=\frac{V_6-V_2}{I}=\frac{V_5-V_3}{I}=\frac{h}{e^2},
\label{eq:R4p}
\end{align}
where $R_{xx}$ only depends on $\beta$, the scattering probability between the voltage contacts, as expected. Notice that $R_{yx}$ remains quantized in our simple model. This is clearly not the case for a real device where $R_{yx}$ starts to deviate from $h/e^2$ with increasing temperature or probe current. However, despite the simplicity of the model, it shows that $R_{xx}$ is more strongly affected by breakdown than $R_{yx}$, in agreement with the measured $I$-$V$ characteristic shown in Fig.~\ref{fig:LB4p}(b).

\subsection{Breakdown in the three-terminal measurement geometry}

\begin{figure}[h]
\centering
\includegraphics[width=10cm, trim={5cm 7cm 5cm 7cm},clip]{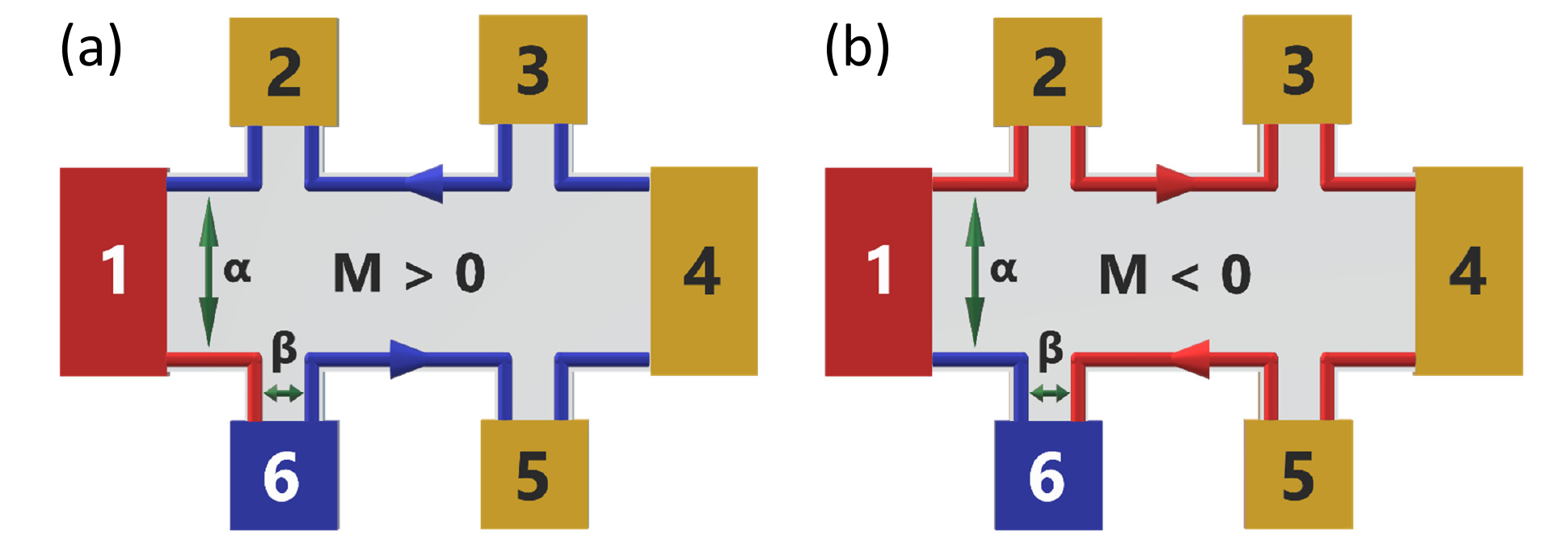}
\caption{(a,b) Schematics of the edge transport in a 6-terminal Hall-bar device for $M>0$ (a) and $M<0$ (b), with the `leakage' between opposing edges for a wide and narrow contacts parametrized by the scattering probabilities $\alpha$ and $\beta$, respectively.
}
\label{fig:LB3p}
\end{figure}

Figure \ref{fig:LB3p} shows a schematic of a 6-terminal Hall-bar device for $M>0$ and $M<0$. The current flows from contacts 1 to 6. Contacts 3, 4 and 5 are spatially separated from the current contacts in this nonlocal configuration. To describe the breakdown in the current contacts, we introduce the scattering probabilities $\alpha$ and $\beta$ between the high- and low-potential branches near contacts 1 and 6, respectively. The nonzero transmission coefficients for $M>0$ then become:
\begin{align}
T_{11}&=\alpha,	&	T_{12}&=(1-\alpha),	&	T_{51}&=\beta(1-\alpha), \nonumber \\
T_{66}&=\beta,	&	T_{56}&=(1-\beta),	&	T_{62}&=\alpha(1-\beta), \\
T_{23}&=T_{34}=T_{45}=1,	&	T_{52}&=\alpha\beta,	&	T_{61}&=(1-\alpha)(1-\beta). \nonumber
\end{align}
The transmission coefficients for $M<0$ are related via $T_{ij}(M<0)=T_{ji}(M>0)$. Using $V_6=0$, $I_1=-I_6=I$ and $I_2=I_3=I_4=I_5=0$, Eq.~\ref{eq:LB} can be solved for $I$ and $V_i$. The three-terminal resistance $R_{1\text{-}6,5\text{-}6}$ then becomes:
\begin{equation}
R_{1\text{-}6,5\text{-}6}=\frac{V_5-V_6}{I}= \begin{cases}
\frac{\beta}{1-\beta}\frac{h}{e^2} & \text{for } M>0,\\
\frac{1}{1-\beta}\frac{h}{e^2} & \text{for } M<0,
\end{cases}
\label{eq:R3p}
\end{equation}
where $R_{1\text{-}6,5\text{-}6}$ only depends on $\beta$, the breakdown in the small contact arm, as one would expect. Moreover, taking the difference $R_{1\text{-}6,5\text{-}6}(M<0)-R_{1\text{-}6,5\text{-}6}(M>0)$ yields $h/e^2$. Hence, the breakdown near the source/drain contact will result in a constant offset in the magnetic field dependence of the three-terminal resistance. Note that all nonlocal resistances are equal, i.e. $R_{1\text{-}6,2\text{-}6}=R_{1\text{-}6,3\text{-}6}=R_{1\text{-}6,4\text{-}6}=R_{1\text{-}6,5\text{-}6}$, as there are no leakage channels in the nonlocal region. If the three-terminal resistance is measured with respect to contact 1, an expression similar to Eq.~\ref{eq:R3p} is found, where the breakdown only depends on $\alpha$ instead of $\beta$. 

Equation \ref{eq:R3p} applies to the three-terminal resistance measurements shown in Fig.~2 of the main text, Fig.~\ref{fig:small3p}, Fig.~\ref{fig:AboveBD}, and Fig.~\ref{fig:contact}. Notice that $V_{25\text{-}0}$ ($I_{1\text{-}0}<0.16$~\si{\uA}) in Fig.~2 and $R_{1\text{-}2,x\text{-}2}$ in Fig.~\ref{fig:small3p}, both measured with respect to a large current contact, correspond to the case of $\beta \approx 0$, i.e. no leakage. On the other hand, all the measurements performed on the 20~\si{\um}-wide contacts of the 26-terminal Hall-bar devices always gave much larger three-terminal resistances than expected for $\beta = 0$ for any value of the probe current. Hence, the breakdown of the QAHE is an issue for small source/drain contacts.

\section{Presence of edge current in the nonlocal region above breakdown}

\begin{figure}
\centering
\includegraphics[width=12cm, trim={0 0 0 4.8cm},clip]{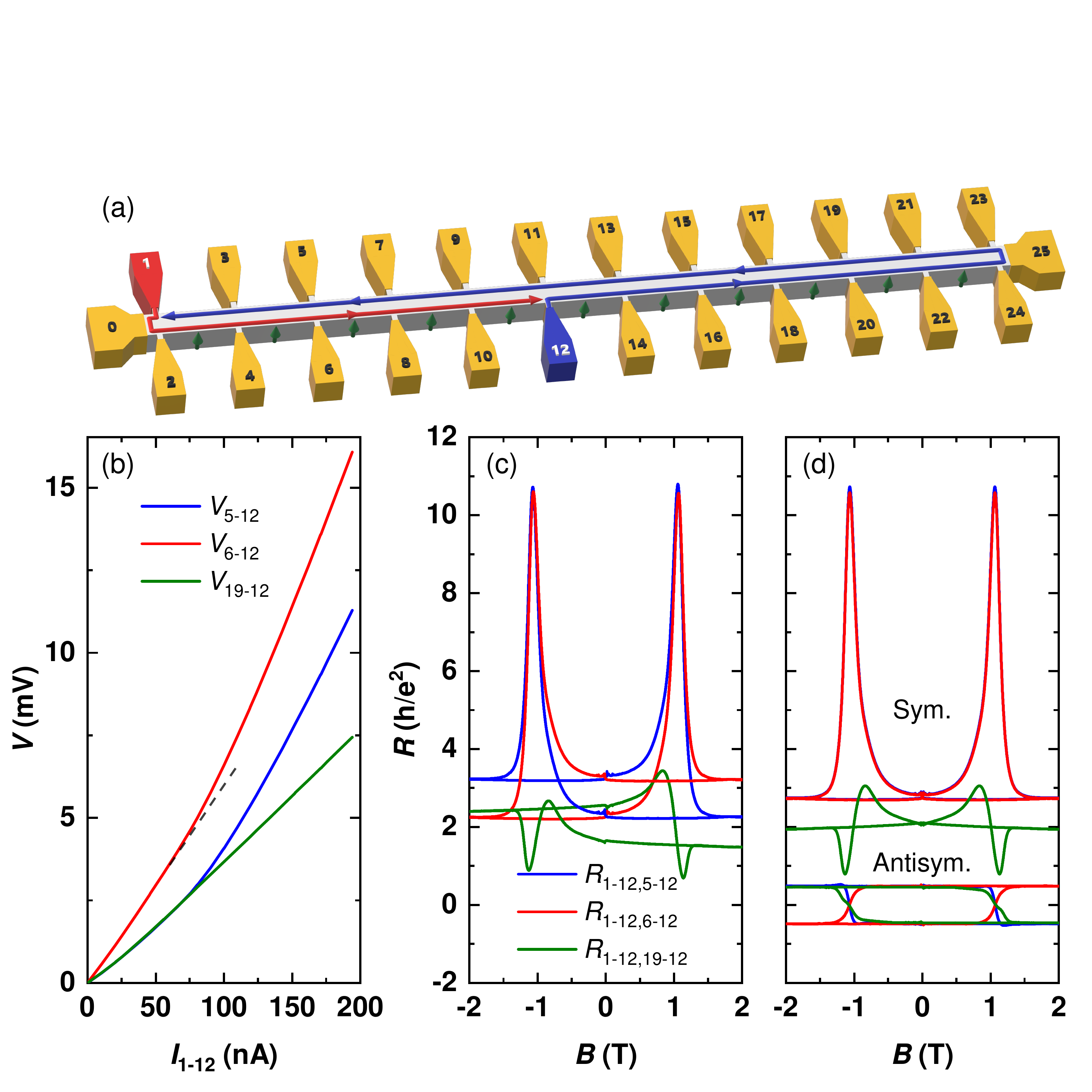}
\caption{Local and nonlocal transport measured above the breakdown current in the 26-terminal Hall-bar device C. (a) Schematic of the high (red) and low (blue) potential portions of the chiral edge state for $M>0$; arrow heads signify the direction of the current flow. The geometry is the same as in Fig.~1 of the main text with the slight difference that contact 12 is grounded instead of contact 14, and that $M$ is reversed. (b) three-terminal voltages $V_{5\text{-}12}$, $V_{6\text{-}12}$, and $V_{19\text{-}12}$ vs the AC current $I_{1\text{-}12}$ measured at +2~T and 20~mK with respect to the 20-\si{\um}-wide contact 12. The breakdown of the QAHE occurs at $\sim$60~nA. (c) Magnetic-field dependencies of the three-terminal resistances $R_{1\text{-}12,5\text{-}12}$ (blue), $R_{1\text{-}12,6\text{-}12}$ (red), and $R_{1\text{-}12,19\text{-}12}$ (green) measured at 20~mK with an AC probe current of 200~nA. (d) Symmetric and antisymmetric components of $R_{1\text{-}12,5\text{-}12}$, $R_{1\text{-}12,6\text{-}12}$, and $R_{1\text{-}12,19\text{-}12}$.
}
\label{fig:AboveBD}
\end{figure}

A possible trivial explanation for the apparent absence of breakdown in the nonlocal region of device A shown in Fig.~1(f) of the main text would be that once breakdown occurs in the local transport region, most or all of the current reaches the drain (contact 14) through dissipative channels rather than the chiral edge state. To address this question, Fig.~\ref{fig:AboveBD}(d) shows the three-terminal $I$-$V$ characteristics measured with respect to the drain contact 12 of device B. The current flows from contacts 1 to 12. For an upward, out-of-plane magnetization ($M>0$), the chiral edge state runs counterclockwise along the sample edge, as shown in Fig.~\ref{fig:AboveBD}(a). Hence, $V_{5\text{-}12}$ and $V_{6\text{-}12}$ correspond to the low- and high-potential branches of the chiral edge state, respectively. Their difference approximately follows the relation $\sim (h/e^2) I_{\text{AC}}$. The breakdown current value for the `local' 100-\si{\um}-wide section of the Hall-bar is $\sim$60~nA, as can be seen from a dent in the curves for $V_{5\text{-}12}$ and $V_{6\text{-}12}$ in Fig.~\ref{fig:AboveBD}(b). Below $\sim$60~nA, $V_{5\text{-}12}$ and $V_{6\text{-}12}$ do not become zero due to a second resistance contribution stemming from the breakdown of the QAHE in the 20-\si{\um}-wide drain contact 12, see Eq.~\ref{eq:R3p}. The voltage $V_{19\text{-}12}$ is nonlocal and corresponds to the low-potential branch of the chiral edge state for $M>0$; as a result, its value is entirely determined by this second resistance contribution stemming from the breakdown in the drain (contact 12). In other words, $V_{5\text{-}12}$ is equal to $V_{19\text{-}12}$ up to $\sim$60~nA where the Hall-bar (with the exception of the source and drain regions) is in the zero resistance state. Above $\sim$60~nA, $V_{5\text{-}12}$ (and $V_{6\text{-}12}$) acquire a larger slope due to the breakdown in the `local' region of the Hall-bar, while the nonlocal voltage $V_{19\text{-}12}$ remains unchanged. These observations speak against the trivial explanation mentioned above.

Figure \ref{fig:AboveBD}(c) shows the magnetic-field dependence of the corresponding three-terminal resistances $R_{1\text{-}12,5\text{-}12}$, $R_{1\text{-}12,6\text{-}12}$, and $R_{1\text{-}12,19\text{-}12}$. In the ideal case (i.e. $\beta$ = 0), a zero to $h/e^2$ transition upon magnetization reversal (cf. Eq.~\ref{eq:R3p}) is expected for $R_{1\text{-}12,5\text{-}12}$, $R_{1\text{-}12,6\text{-}12}$, and $R_{1\text{-}12,19\text{-}12}$. Since the three-terminal resistances are measured with an AC probe current of 200~nA, which is well above the breakdown value, $R_{1\text{-}12,5\text{-}12}$ and $R_{1\text{-}12,6\text{-}12}$ acquire a large longitudinal resistance contribution from the broken-down QAHI state in the `local' region of the Hall-bar. Additionally, $R_{1\text{-}12,5\text{-}12}$, $R_{1\text{-}12,6\text{-}12}$ and $R_{1\text{-}12,19\text{-}12}$ also pick up a second resistance contribution stemming from the breakdown of the QAHE in the 20-\si{\um}-wide drain contact 12, as discussed above.

Figure \ref{fig:AboveBD}(d) shows the symmetric and antisymmetric components of $R_{1\text{-}12,5\text{-}12}$, $R_{1\text{-}12,6\text{-}12}$, and $R_{1\text{-}12,19\text{-}12}$. The symmetric components of $R_{1\text{-}12,5\text{-}12}$ and $R_{1\text{-}12,6\text{-}12}$ are equal to the resistance contribution from the broken-down QAHI state, while their antisymmetric component shows a square hysteresis loop with the resistance value equal to $\sim 0.48h/e^2$. This corresponds to the expected zero to $h/e^2$ transition upon magnetization reversal for this configuration. The deviation of $\sim$4\% from $h/e^2$ is a consequence of the broken QAHI state in the `local' Hall-bar region. Notice that the antisymmetric component of $R_{1\text{-}12,19\text{-}12}$ overlaps with the hysteresis loop of $R_{1\text{-}12,5\text{-}12}$ and $R_{1\text{-}12,6\text{-}12}$, proving the presence of the QAHE edge current in the nonlocal region even above the breakdown. Based on the transport data measured below and above the breakdown (Fig.~1 of the main text and Fig.~\ref{fig:AboveBD}, respectively) it can be concluded that the breakdown of the QAHE solely occurs in the `local' transport region of the Hall-bar device.

\section{Additional data for the three-terminal resistance of device~B}

To gain further insight into the effect of the current-induced breakdown near the source/drain contacts, the magnetic-field dependence of the three-terminal resistance of the 20-\si{\um}-wide contact 2 is shown in Fig.~\ref{fig:contact}. The resistances $R_{1\text{-}2,0\text{-}2}$ and $R_{1\text{-}2,25\text{-}2}$ are measured with contact 2 as the drain. For an upward, out-of-plane magnetization ($M>0$), $R_{1\text{-}2,0\text{-}2}$ represents the high-potential branch of the chiral edge state, while $R_{1\text{-}2,25\text{-}2}$ represents the low-potential branch. In the absence of breakdown, the hysteresis loop of $R_{1\text{-}2,0\text{-}2}$ ($R_{1\text{-}2,25\text{-}2}$) should go from zero ($h/e^2$) at $M<0$ to $h/e^2$ (zero) at $M>0$, see Eq.~\ref{eq:R3p}. For the 20-\si{\um}-wide contact 2, however, the hysteresis loop is offset by about $\sim$31~k$\Omega$. When the difference $\Delta R = R_{1\text{-}2,0\text{-}2} - R_{1\text{-}2,25\text{-}2}$ is taken, both the resistance peaks at the the coercive field and the offset due to breakdown disappear [green curves in Fig.~\ref{fig:contact}(b)]. The $\Delta R$ curves show a clean hysteresis loop from $+h/e^2$ to $-h/e^2$, as expected for a four-terminal Hall measurement of the QAHE below breakdown. This demonstrates that the breakdown affecting the three-terminal measurement is confined to the region with the shortest separation between the high- and low-potential branches of the chiral edge state, i.e. at the source/drain contacts.

\begin{figure}[h]
\centering
\includegraphics[width=14cm]{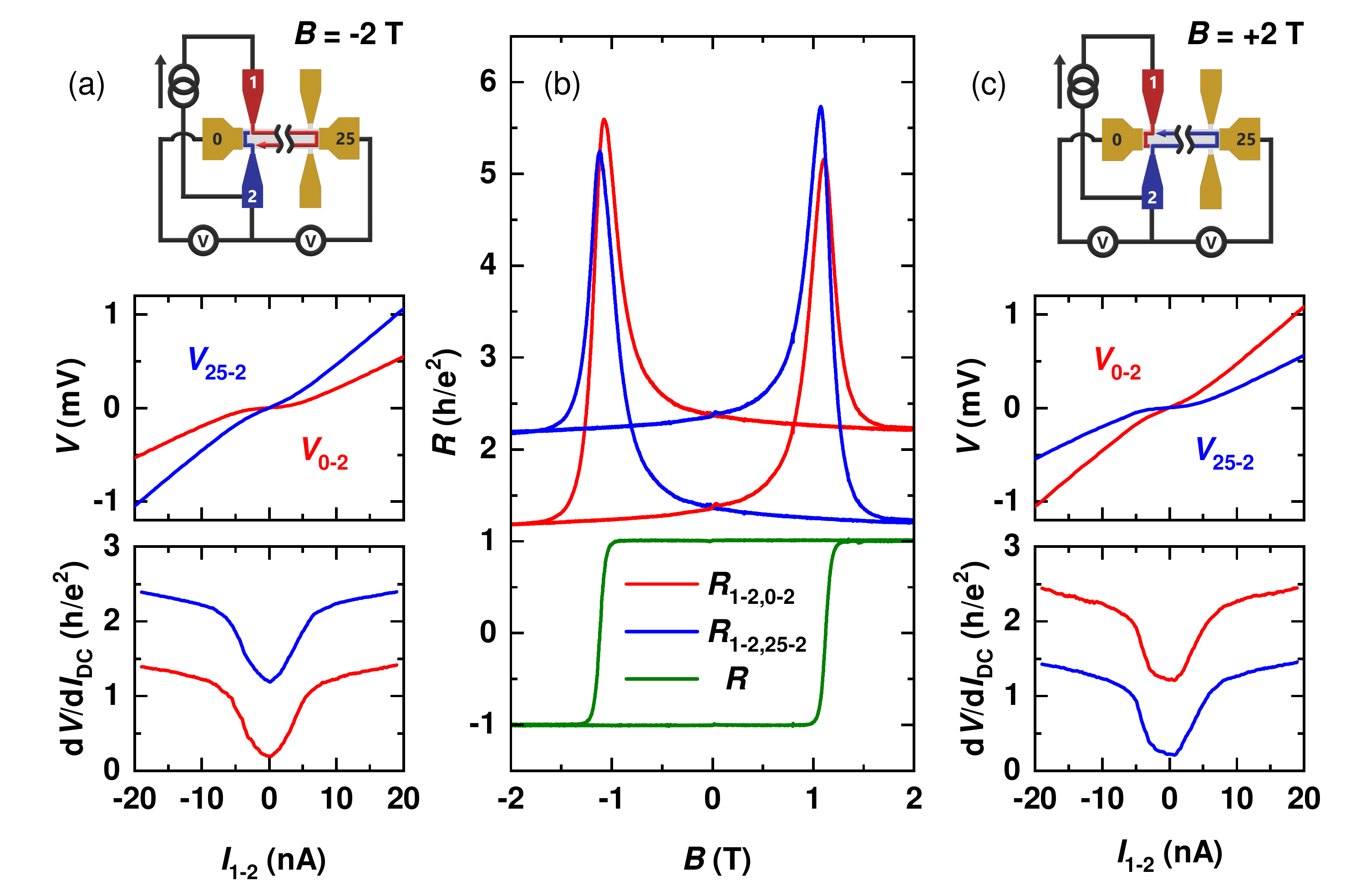}
\caption{The three-terminal resistance of the 20-\si{\um}-wide contact 2 of the 26-terminal Hall-bar device B measured at 40~mK. (a) Schematic of the high (red) and low (blue) potential branch of the chiral edge state at $B=-2$~T with the measurement-circuit configuration (top), corresponding $I$-$V$ characteristics (middle), and the corresponding differential resistances (bottom). (b) Magnetic-field dependencies of the three-terminal resistances $R_{1\text{-}2,0\text{-}2}$ and $R_{1\text{-}2,25\text{-}2}$, measured with $I_{\text{DC}}=30$~nA. $\Delta R \equiv R_{25\text{-}2}-R_{0\text{-}2}$ reconstructs the four-terminal transverse resistance. (c) Schematic of the chiral edge state potential at $B=+2$~T with the measurement-circuit configuration (top), corresponding $I$-$V$ characteristics (middle), and the corresponding differential resistances (bottom). 
}
\label{fig:contact}
\end{figure}

\section{Additional measurement geometries for the 26-terminal Hall-bar device}

\begin{figure}
\centering
\includegraphics[width=\linewidth]{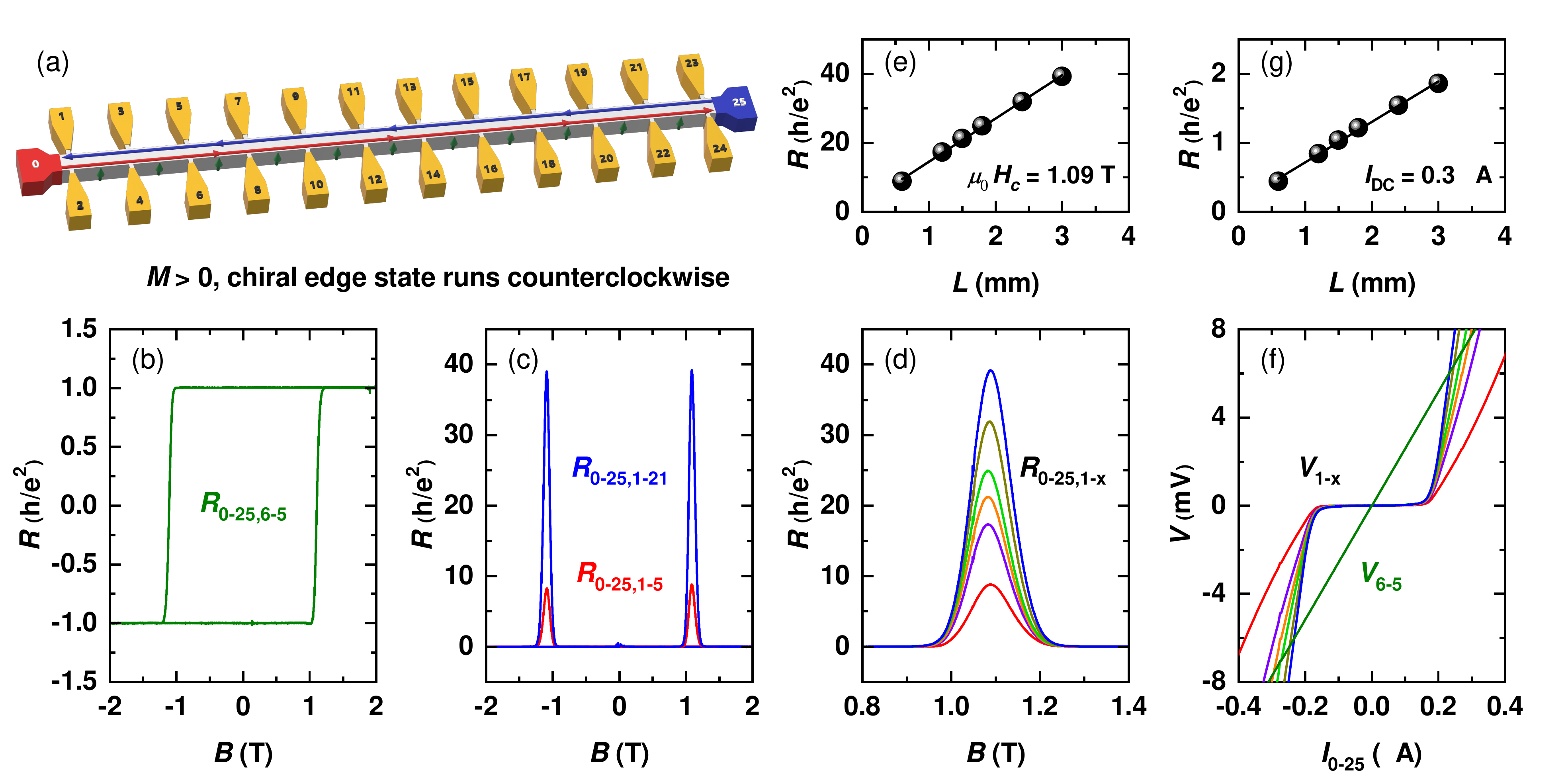}
\caption{`Normal' Hall-bar configuration for the 26-terminal Hall-bar device B, measured at 40~mK with $I_{\text{DC}}=30$~nA. (a) Schematic of the potential distribution in the chiral edge state for $M>0$; red (blue) color signifies the source (drain) potential. (b) Magnetic-field dependence of the transverse resistance $R_{0\text{-}25,6\text{-}5}$. (c) Magnetic-field dependence of the longitudinal resistance $R_{0\text{-}25,1\text{-}5}$ and $R_{0\text{-}25,1\text{-}21}$ with the contact separation of 0.6 and 3~mm, respectively. (d) Longitudinal resistance $R_{0\text{-}25,1\text{-}x}$ with $x = \{ 5,9,11,13,17,21 \}$ near the coercive field $\mu_0 H_c =1.09$~T. (e) The peak values of $R_{0\text{-}25,1\text{-}x}$ at the coercive field $H_c$ as a function the voltage-contact spacing $L$. (f) Plots of $V_{1\text{-}x}$ vs $I_{0\text{-}25}$ measured at +2~T; the transverse voltage $V_{6\text{-}5}$ is also plotted for comparison. The breakdown of the QAHE occurs at $\sim$0.16~\si{\uA}. (g) The value of $V_{1\text{-}x}/I_{0\text{-}25}$ at 0.3~\si{\uA} as a function the voltage contact spacing $L$.
}
\label{fig:normal}
\end{figure}

\begin{figure}
\centering
\includegraphics[width=14cm]{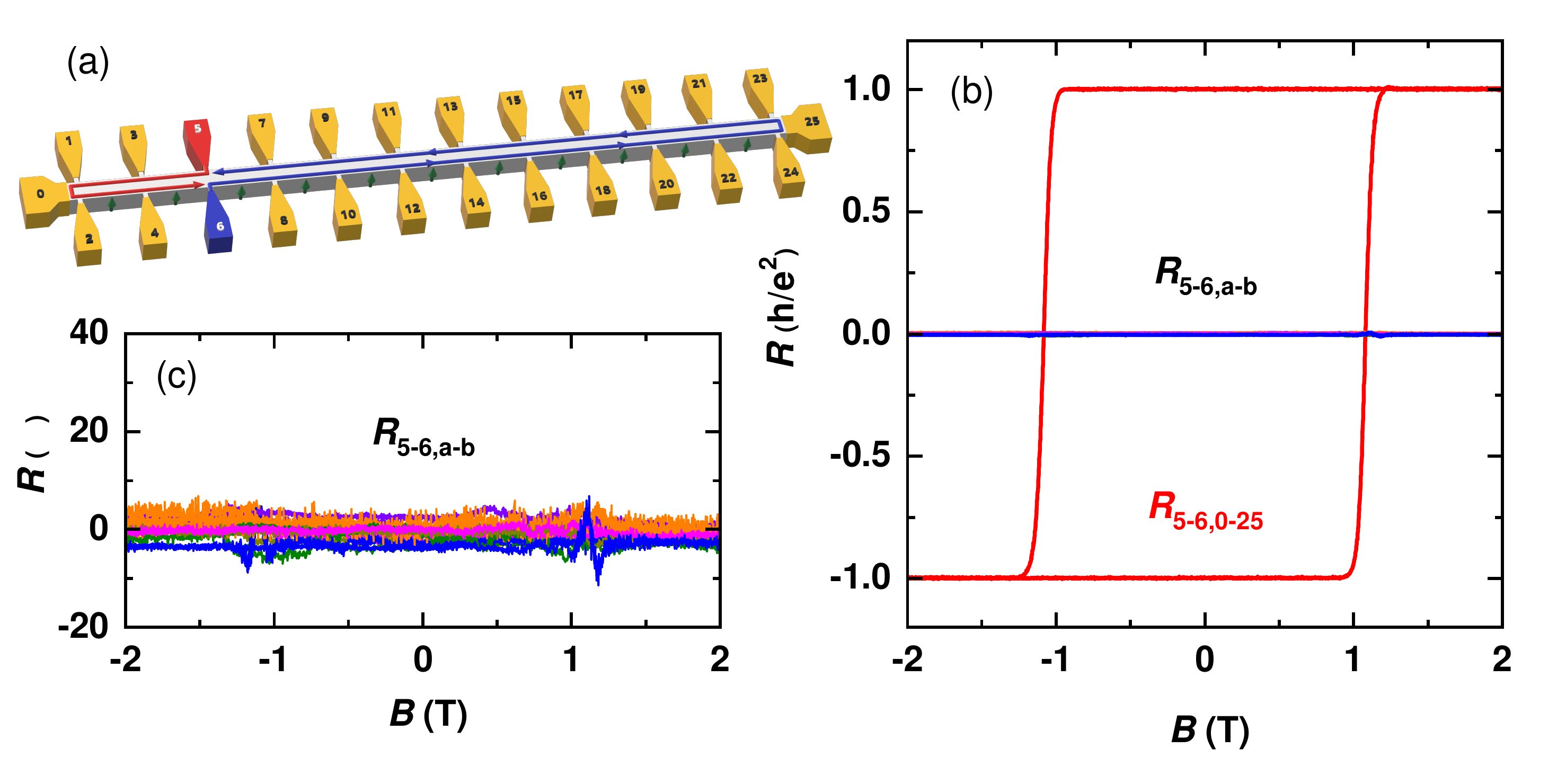}
\caption{`90$^{\circ}$ rotated' Hall-bar configuration for the 26-terminal Hall-bar device A, measured at 30~mK with $I_{\text{DC}}=80$~nA. (a) Schematic of the potential distribution in the chiral edge state for $M>0$. (b) Magnetic-field dependencies of $R_{5\text{-}6,a\text{-}b}$ with $a\text{-}b = \{ 1\text{-}2,9\text{-}10,13\text{-}14,17\text{-}18,21\text{-}22,17\text{-}21 \}$ and the transverse resistance $R_{5\text{-}6,0\text{-}25}$. (c) Magnified plot of $R_{5\text{-}6,a\text{-}b}$ vs $B$.
}
\label{fig:90deg}\bigskip
\includegraphics[width=10cm, trim={0 0 0 3cm},clip]{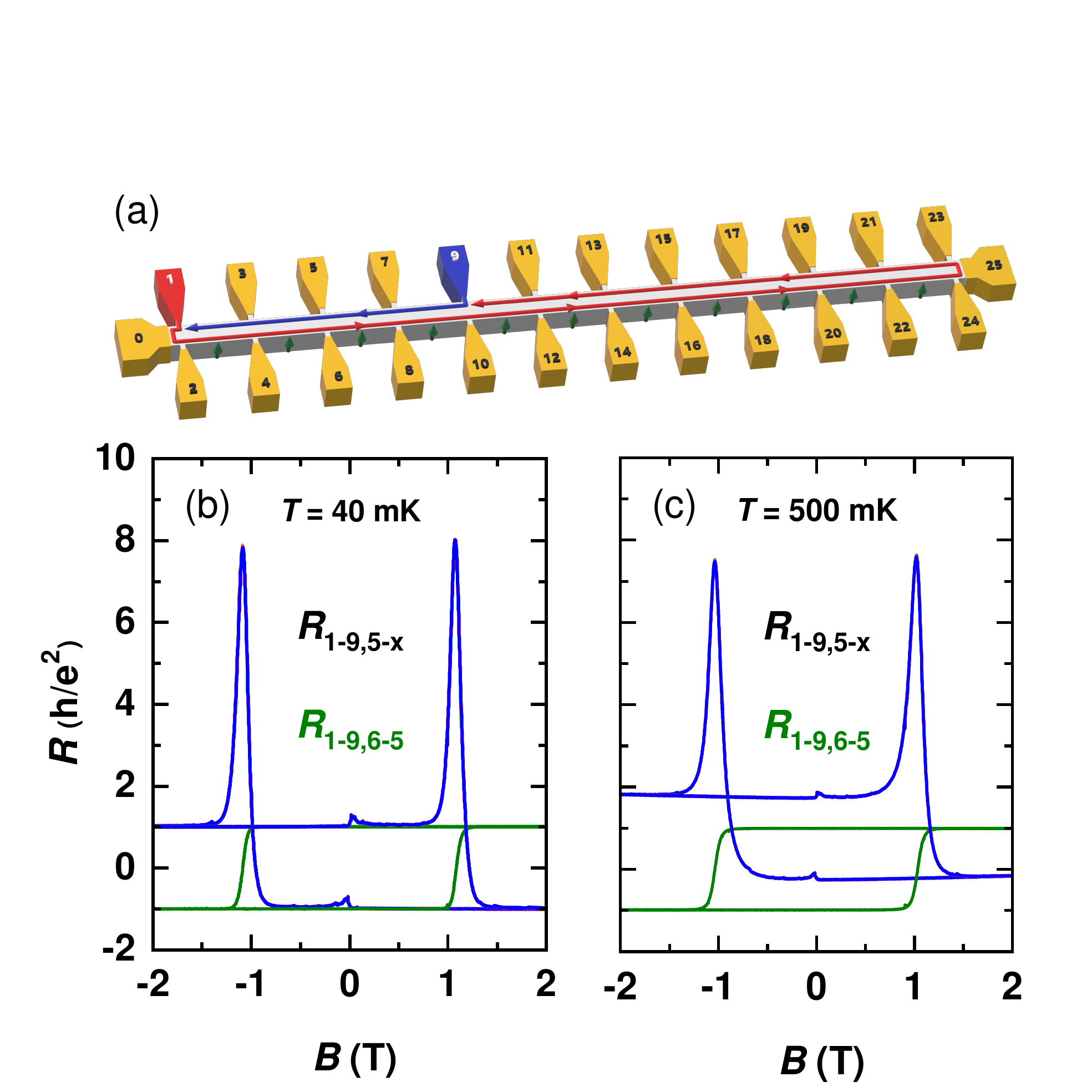}
\caption{`Left edge' configuration for the 26-terminal Hall-bar device A, measured with $I_{\text{DC}}=80$~nA. (a) Schematic of the potential distribution in the chiral edge state for $M>0$. (b,c) Magnetic-field dependencies of $R_{1\text{-}9,5\text{-}x}$ with $x = \{ 13,14,17,18,21,22,25 \}$ and the transverse resistance $R_{1\text{-}9,6\text{-}5}$ measured at 40~mK (b) and at 500~mK (c).
}
\label{fig:LeftEdge}
\end{figure}

\begin{figure}
\centering
\includegraphics[width=7.5cm]{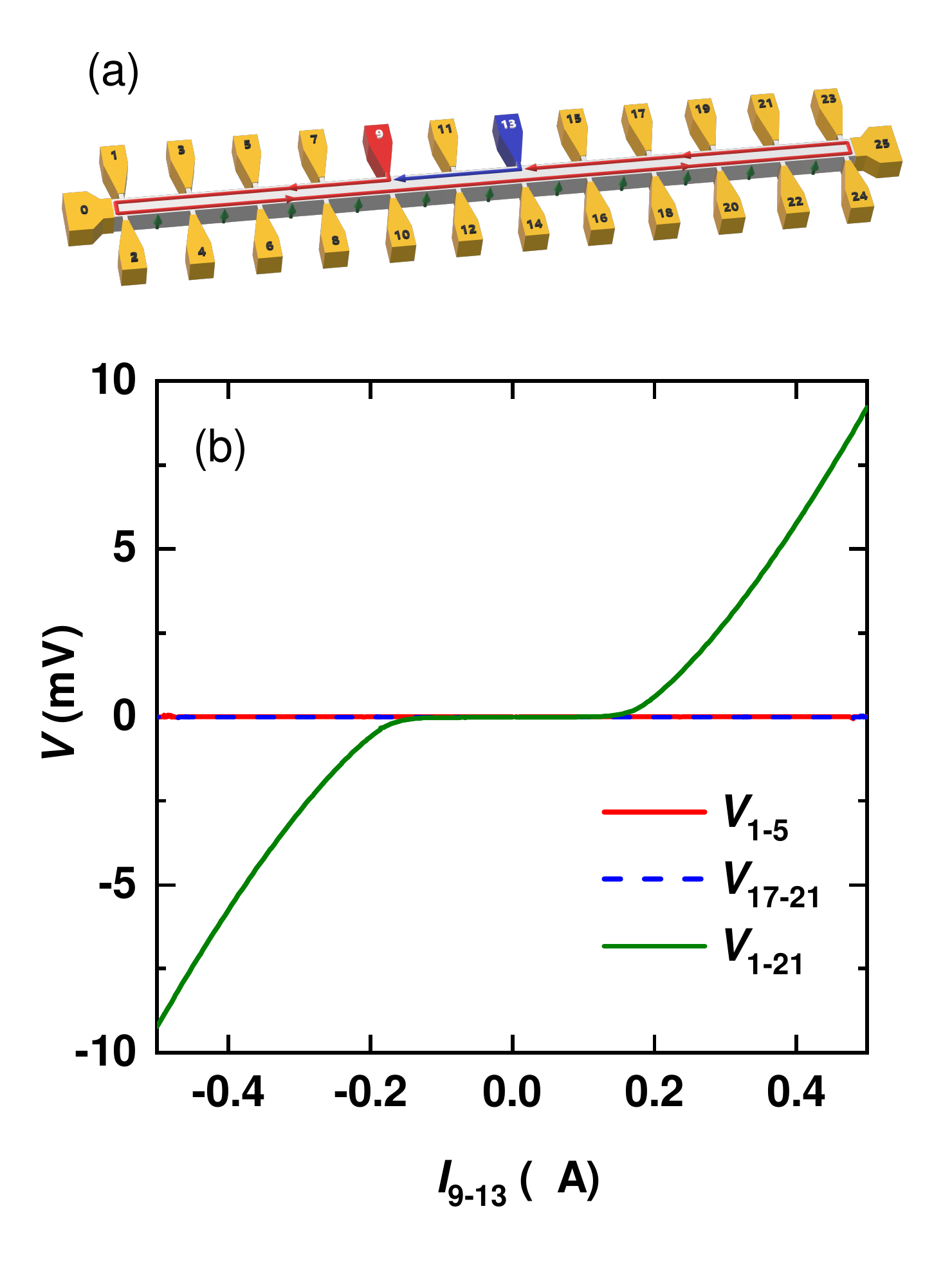}
\caption{`Middle edge' configuration for the 26-terminal Hall-bar device B, measured at 40~mK in +2~T. (a) Schematic of the potential distribution in the chiral edge state for $M>0$.  (b) $I$-$V$ characteristics for $V_{1\text{-}5}$, $V_{17\text{-}21}$, and $V_{1\text{-}21}$ with $I_{9\text{-}13}$. The voltages $V_{1\text{-}5}$ and $V_{17\text{-}21}$ are nonlocal, while $V_{1\text{-}21}$ includes the `local' contribution from the portion between contacts 9 and 13. The breakdown of the QAHE occurs at $\sim$0.16~\si{\uA}.
}
\label{fig:MiddleEdge}
\end{figure}

\noindent In the main text, the long 26-terminal Hall-bar device was used to demonstrate:
\begin{enumerate}
\item Clean nonlocal transport over a mm-size QAHI film.
\item The absence of breakdown of the QAHE in the nonlocal region, identifying the transverse electric field as the driving force of the breakdown.
\item The absence of any dissipation in the nonlocal region, which speaks against the presence of additional quasi-helical edge states as proposed in Ref.~\cite{Wang2013s}.
\end{enumerate}
The measurement geometries in Figs. \ref{fig:normal}, \ref{fig:90deg}, \ref{fig:LeftEdge}, and \ref{fig:MiddleEdge} further support these claims. 
\vspace{2mm}

Figure \ref{fig:normal} shows the `normal' Hall-bar configuration, with the current flowing from contacts 0 to 25. Panels (b-d) show the magnetic-field dependencies of the longitudinal and transverse resistance. Panel (e) shows a linear relation between the peak value of the longitudinal resistance at the coercive field $H_c$ and the voltage contact spacing $L$. Panel (f) shows the current-induced breakdown of the QAHE at $\sim$0.16~\si{\uA}. Panel (g) shows the longitudinal resistance value at $I_{\text{DC}}=0.3$~\si{\uA} as a function of $L$; a linear dependence on the contact spacing is found, similar to that in panel (e), but the slope is smaller by a factor of $\sim$20. This linear relationship is in line with 2D diffusive transport and differs from the exponential length dependence proposed for quasi-helical edge states in Ref. \cite{Wang2013s}.

Figure \ref{fig:90deg} shows the `90$^{\circ}$ rotated' Hall-bar configuration, with the current flowing from contacts 5 to 6. The transverse resistance $R_{5\text{-}6,0\text{-}25}$ is quantized, demonstrating again that the chiral edge state runs along the edge of the sample. Figure \ref{fig:90deg}(c) shows the resistance $R_{5\text{-}6,a\text{-}b}$ with $a\text{-}b$ being the contact pairs $\{ 1\text{-}2,9\text{-}10,13\text{-}14,17\text{-}18,21\text{-}22,17\text{-}21 \}$. $R_{5\text{-}6,a\text{-}b}$ remains essentially zero for any value of the magnetic field, showing near-dissipationless nonlocal edge transport. Moreover, $R_{5\text{-}6,a\text{-}b}$ shows no exponential length dependence as proposed for additional quasi-helical edge states that are not protected from backscattering \cite{Wang2013s}. It should be noted, however, that the large nonlocal device geometry used here might not be best suited to study these proposed dissipative edge states, as their observation might be limited by the phase coherent length $L_\phi$ and the proposed features in Ref.~\cite{Wang2013s} would be strongly suppressed for our large probe separations of 300~\si{\um}.

Figure \ref{fig:LeftEdge} shows the `left edge' Hall-bar configuration, with the current flowing from contacts 1 to 9. The transverse resistance $R_{1\text{-}9,6\text{-}5}$ is quantized at 40 and 500~mK. The resistances $R_{1\text{-}9,5\text{-}x}$, with the voltage contact $x$ being $\{ 13,14,17,18,21,22,25 \}$, demonstrate a quantized hysteresis loop at 40~mK as well, but acquire an additional offset at 500~mK due to a nonzero longitudinal resistance at this elevated temperature. Notice that the hysteresis loops of $R_{1\text{-}9,6\text{-}5}$ and $R_{1\text{-}9,5\text{-}x}$ are opposite, since the polarity of the voltage contacts is switched. Both the offset and the peaks at the coercive field originate from 2D diffusive transport between contacts 1 and 9. $R_{1\text{-}9,5\text{-}x}$ remains the same for all voltage contact $x$ of $\{ 13,14,17,18,21,22,25 \}$, and hence there is no length dependence in the nonlocal region. Moreover, the reported `nonlocal hysteresis loop' (and its temperature dependence) assigned to the existence of additional quasi-helical edge states \cite{Kou2014s,Chang2015Bs} was not observed.

Lastly, Fig.~\ref{fig:MiddleEdge} shows the `middle edge' Hall-bar configuration, with the current flowing from contacts 9 to 13. The current dependence of $V_{1\text{-}5}$ and $V_{17\text{-}21}$, measured nonlocally, shows no sign of breakdown. The voltage $V_{1\text{-}21}$, on the other hand, includes both $V_{1\text{-}5}$ and $V_{17\text{-}21}$ as well as the small `local' region in-between. This $V_{1\text{-}21}$ shows the breakdown of the QAHE at $\sim$0.16~\si{\uA}, demonstrating again that the breakdown only occurs in the `local' transport region.

\section{Comments on other possible breakdown mechanisms for the QAHE}

We first comment on Landau-Zener tunneling between neighboring electron-hole puddles as a possible breakdown mechanism. Estimates of the tunneling probability for such a process in ultrathin TI films possessing a hybridization gap in the 2D surface state spectrum were made in Refs.~\cite{Nandi2018s, Huang2021s}. Assuming a defect density of $N_\text{def} \approx 10^{19}$~cm$^{-3}$, an insulting state in our QAHI films is expected to be realized for a gap $\Delta >$ 10--60 meV at the Dirac point. This is exactly the range of the exchange gap found for magnetically-doped (Bi$_x$Sb$_{1-x}$)$_2$Te$_3$  \cite{Lee2015s, Chong2020s}. Note that in the presence of charge puddles, the application on an electric field is not required to induce Zenner tunneling, because the disorder potential provides the required local electric field and band bending. In other words, in the presence of puddles, Zener tunneling would provide a finite bulk short even at infinitesimally small currents. As a result, no sudden onset of Zenner tunneling at some critical current is expected. 

It is prudent to mention that an estimate based on Refs.~\cite{Nandi2018s, Huang2021s} would predict a sizable Zenner tunneling in our films. However, a near-dissipationless QAHI state has been experimentally observed, at least for low probe currents, with a reported longitudinal resistivity value as low as 1.9~m$\Omega$~\cite{Fox2018s}, which speaks against any major role of Zenner tunneling in compensated QAHIs. Nevertheless, Zenner tunneling might be relevant to the small, non-vanishing resistance in the pre-breakdown regime.

Next we discuss the role of electron heating in the breakdown process. It is obvious that an abrupt increase in the longitudinal resistance will lead to heating, and that the resulting increase of the electron temperature might accelerate the breakdown process. However, the assumption that electron heating itself lies at the origin of the breakdown of the QAHE is doubtful. In the bootstrap electron heating (BSEH) model, initially proposed for the integer quantum Hall effect (QHE) \cite{Komiyama2000s}, the breakdown is attributed to runaway electron heating and described by the following balance equation:
\begin{equation}
\rho_{xx}(T_{el})j^2 = \frac{\epsilon(T_{el})-\epsilon(T_{L})}{\tau_0},
\label{eq:BSEH}
\end{equation}
with $j$ the current density, $\epsilon(T)$ the energy of the system at temperature $T$, $T_{el}$ and $T_{L}$ the electron and lattice temperature, respectively, and $\tau_0$ the relaxation time of the heated electrons \cite{Komiyama2000s}. Upon increasing the current, the energy gained by electrons, $\rho_{xx}(T_{el})j^2$, causes Eq.~\ref{eq:BSEH} to become unstable and a new equilibrium is found at a higher $T_{el}$. The model yields good agreement with the experimentally observed critical field values of the QHE \cite{Nachtwei1999s, Komiyama2000s}. Moreover, it is quite generic and independent of the microscopic details of the samples. Hence, BSEH can be easily employed to describe the breakdown of the QAHE as well \cite{Fox2018s}.
 
However, by comparing the breakdown current normalized by the sample width for both the QHE ($\sim$1~A/m \cite{Jeckelmann2001s}) and the QAHE ($\sim$1~mA/m), it is clear that the heating effect differs by several orders of magnitude. Even if one considers the much smaller excitation energy of the QAHE ($\sim$40~$\mu$eV) compared to the QHE ($\hbar\omega_c \approx 10$~meV at 10~T, with $\omega_c$ the cyclotron frequency), runaway electron heating seems unlikely. Moreover, if one compares the shape of the breakdown curve of the QHE to that of the QAHE, the large vertical jump in the longitudinal resistance at $j_c$ in the QHE case, attributed to the jump in $T_{el}$, is absent in the QAHE breakdown curves. Hence, while electron heating will definitely accelerate the breakdown of the QAHE, it does not seem to be its origin.


%


\begin{thebibliography}{37}%
\makeatletter
\providecommand \@ifxundefined [1]{%
 \@ifx{#1\undefined}
}%
\providecommand \@ifnum [1]{%
 \ifnum #1\expandafter \@firstoftwo
 \else \expandafter \@secondoftwo
 \fi
}%
\providecommand \@ifx [1]{%
 \ifx #1\expandafter \@firstoftwo
 \else \expandafter \@secondoftwo
 \fi
}%
\providecommand \natexlab [1]{#1}%
\providecommand \enquote  [1]{``#1''}%
\providecommand \bibnamefont  [1]{#1}%
\providecommand \bibfnamefont [1]{#1}%
\providecommand \citenamefont [1]{#1}%
\providecommand \href@noop [0]{\@secondoftwo}%
\providecommand \href [0]{\begingroup \@sanitize@url \@href}%
\providecommand \@href[1]{\@@startlink{#1}\@@href}%
\providecommand \@@href[1]{\endgroup#1\@@endlink}%
\providecommand \@sanitize@url [0]{\catcode `\\12\catcode `\$12\catcode
  `\&12\catcode `\#12\catcode `\^12\catcode `\_12\catcode `\%12\relax}%
\providecommand \@@startlink[1]{}%
\providecommand \@@endlink[0]{}%
\providecommand \url  [0]{\begingroup\@sanitize@url \@url }%
\providecommand \@url [1]{\endgroup\@href {#1}{\urlprefix }}%
\providecommand \urlprefix  [0]{URL }%
\providecommand \Eprint [0]{\href }%
\providecommand \doibase [0]{https://doi.org/}%
\providecommand \selectlanguage [0]{\@gobble}%
\providecommand \bibinfo  [0]{\@secondoftwo}%
\providecommand \bibfield  [0]{\@secondoftwo}%
\providecommand \translation [1]{[#1]}%
\providecommand \BibitemOpen [0]{}%
\providecommand \bibitemStop [0]{}%
\providecommand \bibitemNoStop [0]{.\EOS\space}%
\providecommand \EOS [0]{\spacefactor3000\relax}%
\providecommand \BibitemShut  [1]{\csname bibitem#1\endcsname}%
\let\auto@bib@innerbib\@empty
\bibitem [{\citenamefont {Yu}\ \emph {et~al.}(2010)\citenamefont {Yu},
  \citenamefont {Zhang}, \citenamefont {Zhang}, \citenamefont {Zhang},
  \citenamefont {Dai},\ and\ \citenamefont {Fang}}]{Yu2010}%
  \BibitemOpen
  \bibfield  {author} {\bibinfo {author} {\bibfnamefont {R.}~\bibnamefont
  {Yu}}, \bibinfo {author} {\bibfnamefont {W.}~\bibnamefont {Zhang}}, \bibinfo
  {author} {\bibfnamefont {H.-J.}\ \bibnamefont {Zhang}}, \bibinfo {author}
  {\bibfnamefont {S.-C.}\ \bibnamefont {Zhang}}, \bibinfo {author}
  {\bibfnamefont {X.}~\bibnamefont {Dai}},\ and\ \bibinfo {author}
  {\bibfnamefont {Z.}~\bibnamefont {Fang}},\ }\href
  {https://doi.org/10.1126/science.1187485} {\bibfield  {journal} {\bibinfo
  {journal} {Science}\ }\textbf {\bibinfo {volume} {329}},\ \bibinfo {pages}
  {61} (\bibinfo {year} {2010})}\BibitemShut {NoStop}%
\bibitem [{\citenamefont {Chang}\ \emph {et~al.}(2013)\citenamefont {Chang}
  \emph {et~al.}}]{Chang2013}%
  \BibitemOpen
  \bibfield  {author} {\bibinfo {author} {\bibfnamefont {C.-Z.}\ \bibnamefont
  {Chang}} \emph {et~al.},\ }\href {https://doi.org/10.1126/science.1234414}
  {\bibfield  {journal} {\bibinfo  {journal} {Science}\ }\textbf {\bibinfo
  {volume} {340}},\ \bibinfo {pages} {167} (\bibinfo {year}
  {2013})}\BibitemShut {NoStop}%
\bibitem [{\citenamefont {Chang}\ \emph
  {et~al.}(2015{\natexlab{a}})\citenamefont {Chang}, \citenamefont {Zhao},
  \citenamefont {Kim}, \citenamefont {Zhang}, \citenamefont {Assaf},
  \citenamefont {Heiman}, \citenamefont {Zhang}, \citenamefont {Liu},
  \citenamefont {Chan},\ and\ \citenamefont {Moodera}}]{Chang2015A}%
  \BibitemOpen
  \bibfield  {author} {\bibinfo {author} {\bibfnamefont {C.-Z.}\ \bibnamefont
  {Chang}}, \bibinfo {author} {\bibfnamefont {W.}~\bibnamefont {Zhao}},
  \bibinfo {author} {\bibfnamefont {D.~Y.}\ \bibnamefont {Kim}}, \bibinfo
  {author} {\bibfnamefont {H.}~\bibnamefont {Zhang}}, \bibinfo {author}
  {\bibfnamefont {B.~A.}\ \bibnamefont {Assaf}}, \bibinfo {author}
  {\bibfnamefont {D.}~\bibnamefont {Heiman}}, \bibinfo {author} {\bibfnamefont
  {S.-C.}\ \bibnamefont {Zhang}}, \bibinfo {author} {\bibfnamefont
  {C.}~\bibnamefont {Liu}}, \bibinfo {author} {\bibfnamefont {M.~H.~W.}\
  \bibnamefont {Chan}},\ and\ \bibinfo {author} {\bibfnamefont {J.~S.}\
  \bibnamefont {Moodera}},\ }\href {https://doi.org/10.1038/nmat4204}
  {\bibfield  {journal} {\bibinfo  {journal} {Nat. Mater.}\ }\textbf {\bibinfo
  {volume} {14}},\ \bibinfo {pages} {473} (\bibinfo {year}
  {2015}{\natexlab{a}})}\BibitemShut {NoStop}%
\bibitem [{\citenamefont {{Breunig}}\ and\ \citenamefont
  {{Ando}}(2021)}]{Breunig2021}%
  \BibitemOpen
  \bibfield  {author} {\bibinfo {author} {\bibfnamefont {O.}~\bibnamefont
  {{Breunig}}}\ and\ \bibinfo {author} {\bibfnamefont {Y.}~\bibnamefont
  {{Ando}}},\ }\href@noop {} {\bibfield  {journal} {\bibinfo  {journal} {arXiv
  e-prints}\ ,\ \bibinfo {eid} {arXiv:2101.12538}} (\bibinfo {year} {2021})},\
  \Eprint {https://arxiv.org/abs/2101.12538} {arXiv:2101.12538} \BibitemShut
  {NoStop}%
\bibitem [{\citenamefont {Qi}\ \emph {et~al.}(2008)\citenamefont {Qi},
  \citenamefont {Hughes},\ and\ \citenamefont {Zhang}}]{Qi2008}%
  \BibitemOpen
  \bibfield  {author} {\bibinfo {author} {\bibfnamefont {X.-L.}\ \bibnamefont
  {Qi}}, \bibinfo {author} {\bibfnamefont {T.~L.}\ \bibnamefont {Hughes}},\
  and\ \bibinfo {author} {\bibfnamefont {S.-C.}\ \bibnamefont {Zhang}},\ }\href
  {https://doi.org/10.1103/PhysRevB.78.195424} {\bibfield  {journal} {\bibinfo
  {journal} {Phys. Rev. B}\ }\textbf {\bibinfo {volume} {78}},\ \bibinfo
  {pages} {195424} (\bibinfo {year} {2008})}\BibitemShut {NoStop}%
\bibitem [{\citenamefont {Qi}\ \emph {et~al.}(2010)\citenamefont {Qi},
  \citenamefont {Hughes},\ and\ \citenamefont {Zhang}}]{Qi2010}%
  \BibitemOpen
  \bibfield  {author} {\bibinfo {author} {\bibfnamefont {X.-L.}\ \bibnamefont
  {Qi}}, \bibinfo {author} {\bibfnamefont {T.~L.}\ \bibnamefont {Hughes}},\
  and\ \bibinfo {author} {\bibfnamefont {S.-C.}\ \bibnamefont {Zhang}},\ }\href
  {https://doi.org/10.1103/PhysRevB.82.184516} {\bibfield  {journal} {\bibinfo
  {journal} {Phys. Rev. B}\ }\textbf {\bibinfo {volume} {82}},\ \bibinfo
  {pages} {184516} (\bibinfo {year} {2010})}\BibitemShut {NoStop}%
\bibitem [{\citenamefont {Kou}\ \emph {et~al.}(2014)\citenamefont {Kou} \emph
  {et~al.}}]{Kou2014}%
  \BibitemOpen
  \bibfield  {author} {\bibinfo {author} {\bibfnamefont {X.}~\bibnamefont
  {Kou}} \emph {et~al.},\ }\href
  {https://doi.org/10.1103/PhysRevLett.113.137201} {\bibfield  {journal}
  {\bibinfo  {journal} {Phys. Rev. Lett.}\ }\textbf {\bibinfo {volume} {113}},\
  \bibinfo {pages} {137201} (\bibinfo {year} {2014})}\BibitemShut {NoStop}%
\bibitem [{\citenamefont {Chang}\ \emph
  {et~al.}(2015{\natexlab{b}})\citenamefont {Chang}, \citenamefont {Zhao},
  \citenamefont {Kim}, \citenamefont {Wei}, \citenamefont {Jain}, \citenamefont
  {Liu}, \citenamefont {Chan},\ and\ \citenamefont {Moodera}}]{Chang2015B}%
  \BibitemOpen
  \bibfield  {author} {\bibinfo {author} {\bibfnamefont {C.-Z.}\ \bibnamefont
  {Chang}}, \bibinfo {author} {\bibfnamefont {W.}~\bibnamefont {Zhao}},
  \bibinfo {author} {\bibfnamefont {D.~Y.}\ \bibnamefont {Kim}}, \bibinfo
  {author} {\bibfnamefont {P.}~\bibnamefont {Wei}}, \bibinfo {author}
  {\bibfnamefont {J.~K.}\ \bibnamefont {Jain}}, \bibinfo {author}
  {\bibfnamefont {C.}~\bibnamefont {Liu}}, \bibinfo {author} {\bibfnamefont
  {M.~H.~W.}\ \bibnamefont {Chan}},\ and\ \bibinfo {author} {\bibfnamefont
  {J.~S.}\ \bibnamefont {Moodera}},\ }\href
  {https://doi.org/10.1103/PhysRevLett.115.057206} {\bibfield  {journal}
  {\bibinfo  {journal} {Phys. Rev. Lett.}\ }\textbf {\bibinfo {volume} {115}},\
  \bibinfo {pages} {057206} (\bibinfo {year} {2015}{\natexlab{b}})}\BibitemShut
  {NoStop}%
\bibitem [{\citenamefont {Wang}\ \emph {et~al.}(2013)\citenamefont {Wang},
  \citenamefont {Lian}, \citenamefont {Zhang},\ and\ \citenamefont
  {Zhang}}]{Wang2013}%
  \BibitemOpen
  \bibfield  {author} {\bibinfo {author} {\bibfnamefont {J.}~\bibnamefont
  {Wang}}, \bibinfo {author} {\bibfnamefont {B.}~\bibnamefont {Lian}}, \bibinfo
  {author} {\bibfnamefont {H.}~\bibnamefont {Zhang}},\ and\ \bibinfo {author}
  {\bibfnamefont {S.-C.}\ \bibnamefont {Zhang}},\ }\href
  {https://doi.org/10.1103/PhysRevLett.111.086803} {\bibfield  {journal}
  {\bibinfo  {journal} {Phys. Rev. Lett.}\ }\textbf {\bibinfo {volume} {111}},\
  \bibinfo {pages} {086803} (\bibinfo {year} {2013})}\BibitemShut {NoStop}%
\bibitem [{\citenamefont {Kawamura}\ \emph {et~al.}(2017)\citenamefont
  {Kawamura}, \citenamefont {Yoshimi}, \citenamefont {Tsukazaki}, \citenamefont
  {Takahashi}, \citenamefont {Kawasaki},\ and\ \citenamefont
  {Tokura}}]{Kawamura2017}%
  \BibitemOpen
  \bibfield  {author} {\bibinfo {author} {\bibfnamefont {M.}~\bibnamefont
  {Kawamura}}, \bibinfo {author} {\bibfnamefont {R.}~\bibnamefont {Yoshimi}},
  \bibinfo {author} {\bibfnamefont {A.}~\bibnamefont {Tsukazaki}}, \bibinfo
  {author} {\bibfnamefont {K.~S.}\ \bibnamefont {Takahashi}}, \bibinfo {author}
  {\bibfnamefont {M.}~\bibnamefont {Kawasaki}},\ and\ \bibinfo {author}
  {\bibfnamefont {Y.}~\bibnamefont {Tokura}},\ }\href
  {https://doi.org/10.1103/PhysRevLett.119.016803} {\bibfield  {journal}
  {\bibinfo  {journal} {Phys. Rev. Lett.}\ }\textbf {\bibinfo {volume} {119}},\
  \bibinfo {pages} {016803} (\bibinfo {year} {2017})}\BibitemShut {NoStop}%
\bibitem [{\citenamefont {Fox}\ \emph {et~al.}(2018)\citenamefont {Fox},
  \citenamefont {Rosen}, \citenamefont {Yang}, \citenamefont {Jones},
  \citenamefont {Elmquist}, \citenamefont {Kou}, \citenamefont {Pan},
  \citenamefont {Wang},\ and\ \citenamefont {Goldhaber-Gordon}}]{Fox2018}%
  \BibitemOpen
  \bibfield  {author} {\bibinfo {author} {\bibfnamefont {E.~J.}\ \bibnamefont
  {Fox}}, \bibinfo {author} {\bibfnamefont {I.~T.}\ \bibnamefont {Rosen}},
  \bibinfo {author} {\bibfnamefont {Y.}~\bibnamefont {Yang}}, \bibinfo {author}
  {\bibfnamefont {G.~R.}\ \bibnamefont {Jones}}, \bibinfo {author}
  {\bibfnamefont {R.~E.}\ \bibnamefont {Elmquist}}, \bibinfo {author}
  {\bibfnamefont {X.}~\bibnamefont {Kou}}, \bibinfo {author} {\bibfnamefont
  {L.}~\bibnamefont {Pan}}, \bibinfo {author} {\bibfnamefont {K.~L.}\
  \bibnamefont {Wang}},\ and\ \bibinfo {author} {\bibfnamefont
  {D.}~\bibnamefont {Goldhaber-Gordon}},\ }\href
  {https://doi.org/10.1103/PhysRevB.98.075145} {\bibfield  {journal} {\bibinfo
  {journal} {Phys. Rev. B}\ }\textbf {\bibinfo {volume} {98}},\ \bibinfo
  {pages} {075145} (\bibinfo {year} {2018})}\BibitemShut {NoStop}%
\bibitem [{\citenamefont {Götz}\ \emph {et~al.}(2018)\citenamefont {Götz},
  \citenamefont {Fijalkowski}, \citenamefont {Pesel}, \citenamefont {Hartl},
  \citenamefont {Schreyeck}, \citenamefont {Winnerlein}, \citenamefont
  {Grauer}, \citenamefont {Scherer}, \citenamefont {Brunner}, \citenamefont
  {Gould}, \citenamefont {Ahlers},\ and\ \citenamefont {Molenkamp}}]{Gotz2018}%
  \BibitemOpen
  \bibfield  {author} {\bibinfo {author} {\bibfnamefont {M.}~\bibnamefont
  {Götz}}, \bibinfo {author} {\bibfnamefont {K.~M.}\ \bibnamefont
  {Fijalkowski}}, \bibinfo {author} {\bibfnamefont {E.}~\bibnamefont {Pesel}},
  \bibinfo {author} {\bibfnamefont {M.}~\bibnamefont {Hartl}}, \bibinfo
  {author} {\bibfnamefont {S.}~\bibnamefont {Schreyeck}}, \bibinfo {author}
  {\bibfnamefont {M.}~\bibnamefont {Winnerlein}}, \bibinfo {author}
  {\bibfnamefont {S.}~\bibnamefont {Grauer}}, \bibinfo {author} {\bibfnamefont
  {H.}~\bibnamefont {Scherer}}, \bibinfo {author} {\bibfnamefont
  {K.}~\bibnamefont {Brunner}}, \bibinfo {author} {\bibfnamefont
  {C.}~\bibnamefont {Gould}}, \bibinfo {author} {\bibfnamefont {F.~J.}\
  \bibnamefont {Ahlers}},\ and\ \bibinfo {author} {\bibfnamefont {L.~W.}\
  \bibnamefont {Molenkamp}},\ }\href {https://doi.org/10.1063/1.5009718}
  {\bibfield  {journal} {\bibinfo  {journal} {Appl. Phys. Lett.}\ }\textbf
  {\bibinfo {volume} {112}},\ \bibinfo {pages} {072102} (\bibinfo {year}
  {2018})}\BibitemShut {NoStop}%
\bibitem [{\citenamefont {Okazaki}\ \emph {et~al.}(2020)\citenamefont
  {Okazaki}, \citenamefont {Oe}, \citenamefont {Kawamura}, \citenamefont
  {Yoshimi}, \citenamefont {Nakamura}, \citenamefont {Takada}, \citenamefont
  {Mogi}, \citenamefont {Takahashi}, \citenamefont {Tsukazaki}, \citenamefont
  {Kawasaki}, \citenamefont {Tokura},\ and\ \citenamefont
  {Kaneko}}]{Okazaki2020}%
  \BibitemOpen
  \bibfield  {author} {\bibinfo {author} {\bibfnamefont {Y.}~\bibnamefont
  {Okazaki}}, \bibinfo {author} {\bibfnamefont {T.}~\bibnamefont {Oe}},
  \bibinfo {author} {\bibfnamefont {M.}~\bibnamefont {Kawamura}}, \bibinfo
  {author} {\bibfnamefont {R.}~\bibnamefont {Yoshimi}}, \bibinfo {author}
  {\bibfnamefont {S.}~\bibnamefont {Nakamura}}, \bibinfo {author}
  {\bibfnamefont {S.}~\bibnamefont {Takada}}, \bibinfo {author} {\bibfnamefont
  {M.}~\bibnamefont {Mogi}}, \bibinfo {author} {\bibfnamefont {K.~S.}\
  \bibnamefont {Takahashi}}, \bibinfo {author} {\bibfnamefont {A.}~\bibnamefont
  {Tsukazaki}}, \bibinfo {author} {\bibfnamefont {M.}~\bibnamefont {Kawasaki}},
  \bibinfo {author} {\bibfnamefont {Y.}~\bibnamefont {Tokura}},\ and\ \bibinfo
  {author} {\bibfnamefont {N.-H.}\ \bibnamefont {Kaneko}},\ }\href
  {https://doi.org/10.1063/1.5145172} {\bibfield  {journal} {\bibinfo
  {journal} {Appl. Phys. Lett.}\ }\textbf {\bibinfo {volume} {116}},\ \bibinfo
  {pages} {143101} (\bibinfo {year} {2020})}\BibitemShut {NoStop}%
\bibitem [{\citenamefont {Beenakker}\ \emph {et~al.}(2019)\citenamefont
  {Beenakker}, \citenamefont {Baireuther}, \citenamefont {Herasymenko},
  \citenamefont {Adagideli}, \citenamefont {Wang},\ and\ \citenamefont
  {Akhmerov}}]{Beenakker2019}%
  \BibitemOpen
  \bibfield  {author} {\bibinfo {author} {\bibfnamefont {C.~W.~J.}\
  \bibnamefont {Beenakker}}, \bibinfo {author} {\bibfnamefont {P.}~\bibnamefont
  {Baireuther}}, \bibinfo {author} {\bibfnamefont {Y.}~\bibnamefont
  {Herasymenko}}, \bibinfo {author} {\bibfnamefont {I.}~\bibnamefont
  {Adagideli}}, \bibinfo {author} {\bibfnamefont {L.}~\bibnamefont {Wang}},\
  and\ \bibinfo {author} {\bibfnamefont {A.~R.}\ \bibnamefont {Akhmerov}},\
  }\href {https://doi.org/10.1103/PhysRevLett.122.146803} {\bibfield  {journal}
  {\bibinfo  {journal} {Phys. Rev. Lett.}\ }\textbf {\bibinfo {volume} {122}},\
  \bibinfo {pages} {146803} (\bibinfo {year} {2019})}\BibitemShut {NoStop}%
\bibitem [{\citenamefont {Adagideli}\ \emph {et~al.}(2020)\citenamefont
  {Adagideli}, \citenamefont {Hassler}, \citenamefont {A.}, \citenamefont
  {Pacholski},\ and\ \citenamefont {Beenakker}}]{Adagideli2020}%
  \BibitemOpen
  \bibfield  {author} {\bibinfo {author} {\bibfnamefont {I.}~\bibnamefont
  {Adagideli}}, \bibinfo {author} {\bibfnamefont {F.}~\bibnamefont {Hassler}},
  \bibinfo {author} {\bibfnamefont {G.}~\bibnamefont {A.}}, \bibinfo {author}
  {\bibfnamefont {M.}~\bibnamefont {Pacholski}},\ and\ \bibinfo {author}
  {\bibfnamefont {C.}~\bibnamefont {Beenakker}},\ }\href
  {https://doi.org/10.21468/SciPostPhys.8.1.013} {\bibfield  {journal}
  {\bibinfo  {journal} {SciPost Phys.}\ }\textbf {\bibinfo {volume} {8}},\
  \bibinfo {pages} {13} (\bibinfo {year} {2020})}\BibitemShut {NoStop}%
\bibitem [{\citenamefont {Hassler}\ \emph {et~al.}(2020)\citenamefont
  {Hassler}, \citenamefont {Grabsch}, \citenamefont {Pacholski}, \citenamefont
  {Oriekhov}, \citenamefont {Ovdat}, \citenamefont {Adagideli},\ and\
  \citenamefont {Beenakker}}]{Hassler2020}%
  \BibitemOpen
  \bibfield  {author} {\bibinfo {author} {\bibfnamefont {F.}~\bibnamefont
  {Hassler}}, \bibinfo {author} {\bibfnamefont {A.}~\bibnamefont {Grabsch}},
  \bibinfo {author} {\bibfnamefont {M.~J.}\ \bibnamefont {Pacholski}}, \bibinfo
  {author} {\bibfnamefont {D.~O.}\ \bibnamefont {Oriekhov}}, \bibinfo {author}
  {\bibfnamefont {O.}~\bibnamefont {Ovdat}}, \bibinfo {author} {\bibfnamefont
  {I.}~\bibnamefont {Adagideli}},\ and\ \bibinfo {author} {\bibfnamefont
  {C.~W.~J.}\ \bibnamefont {Beenakker}},\ }\href
  {https://doi.org/10.1103/PhysRevB.102.045431} {\bibfield  {journal} {\bibinfo
   {journal} {Phys. Rev. B}\ }\textbf {\bibinfo {volume} {102}},\ \bibinfo
  {pages} {045431} (\bibinfo {year} {2020})}\BibitemShut {NoStop}%
\bibitem [{SM()}]{SM}%
  \BibitemOpen
  \href@noop {} {\bibinfo  {journal} {See Supplemental Material at [URL will be
  inserted by publisher] for additional measurement data and discussion}\
  }\BibitemShut {NoStop}%
\bibitem [{\citenamefont {B\"uttiker}(1988)}]{Buttiker1988}%
  \BibitemOpen
\bibfield  {journal} {  }\bibfield  {author} {\bibinfo {author} {\bibfnamefont
  {M.}~\bibnamefont {B\"uttiker}},\ }\href
  {https://doi.org/10.1103/PhysRevB.38.9375} {\bibfield  {journal} {\bibinfo
  {journal} {Phys. Rev. B}\ }\textbf {\bibinfo {volume} {38}},\ \bibinfo
  {pages} {9375} (\bibinfo {year} {1988})}\BibitemShut {NoStop}%
\bibitem [{\citenamefont {Kayyalha}\ \emph {et~al.}(2020)\citenamefont
  {Kayyalha} \emph {et~al.}}]{Kayyalha2020}%
  \BibitemOpen
  \bibfield  {author} {\bibinfo {author} {\bibfnamefont {M.}~\bibnamefont
  {Kayyalha}} \emph {et~al.},\ }\href {https://doi.org/10.1126/science.aax6361}
  {\bibfield  {journal} {\bibinfo  {journal} {Science}\ }\textbf {\bibinfo
  {volume} {367}},\ \bibinfo {pages} {64} (\bibinfo {year} {2020})}\BibitemShut
  {NoStop}%
\bibitem [{\citenamefont {Bestwick}\ \emph {et~al.}(2015)\citenamefont
  {Bestwick}, \citenamefont {Fox}, \citenamefont {Kou}, \citenamefont {Pan},
  \citenamefont {Wang},\ and\ \citenamefont {Goldhaber-Gordon}}]{Bestwick2015}%
  \BibitemOpen
  \bibfield  {author} {\bibinfo {author} {\bibfnamefont {A.~J.}\ \bibnamefont
  {Bestwick}}, \bibinfo {author} {\bibfnamefont {E.~J.}\ \bibnamefont {Fox}},
  \bibinfo {author} {\bibfnamefont {X.}~\bibnamefont {Kou}}, \bibinfo {author}
  {\bibfnamefont {L.}~\bibnamefont {Pan}}, \bibinfo {author} {\bibfnamefont
  {K.~L.}\ \bibnamefont {Wang}},\ and\ \bibinfo {author} {\bibfnamefont
  {D.}~\bibnamefont {Goldhaber-Gordon}},\ }\href
  {https://doi.org/10.1103/PhysRevLett.114.187201} {\bibfield  {journal}
  {\bibinfo  {journal} {Phys. Rev. Lett.}\ }\textbf {\bibinfo {volume} {114}},\
  \bibinfo {pages} {187201} (\bibinfo {year} {2015})}\BibitemShut {NoStop}%
\bibitem [{\citenamefont {Ou}\ \emph {et~al.}(2018)\citenamefont {Ou} \emph
  {et~al.}}]{Ou2017}%
  \BibitemOpen
  \bibfield  {author} {\bibinfo {author} {\bibfnamefont {Y.}~\bibnamefont {Ou}}
  \emph {et~al.},\ }\href {https://doi.org/10.1002/adma.201703062} {\bibfield
  {journal} {\bibinfo  {journal} {Adv. Mater.}\ }\textbf {\bibinfo {volume}
  {30}},\ \bibinfo {pages} {1703062} (\bibinfo {year} {2018})}\BibitemShut
  {NoStop}%
\bibitem [{\citenamefont {Rosen}\ \emph {et~al.}(2017)\citenamefont {Rosen},
  \citenamefont {Fox}, \citenamefont {Kou}, \citenamefont {Pan}, \citenamefont
  {Wang},\ and\ \citenamefont {Goldhaber-Gordon}}]{Rosen2017}%
  \BibitemOpen
  \bibfield  {author} {\bibinfo {author} {\bibfnamefont {I.~T.}\ \bibnamefont
  {Rosen}}, \bibinfo {author} {\bibfnamefont {E.~J.}\ \bibnamefont {Fox}},
  \bibinfo {author} {\bibfnamefont {X.}~\bibnamefont {Kou}}, \bibinfo {author}
  {\bibfnamefont {L.}~\bibnamefont {Pan}}, \bibinfo {author} {\bibfnamefont
  {K.~L.}\ \bibnamefont {Wang}},\ and\ \bibinfo {author} {\bibfnamefont
  {D.}~\bibnamefont {Goldhaber-Gordon}},\ }\href
  {https://doi.org/10.1038/s41535-017-0073-0} {\bibfield  {journal} {\bibinfo
  {journal} {npj Quantum Mater.}\ }\textbf {\bibinfo {volume} {2}},\ \bibinfo
  {pages} {69} (\bibinfo {year} {2017})}\BibitemShut {NoStop}%
\bibitem [{\citenamefont {Lee}\ \emph {et~al.}(2015)\citenamefont {Lee},
  \citenamefont {Kim}, \citenamefont {Lee}, \citenamefont {Billinge},
  \citenamefont {Zhong}, \citenamefont {Schneeloch}, \citenamefont {Liu},
  \citenamefont {Valla}, \citenamefont {Tranquada}, \citenamefont {Gu},\ and\
  \citenamefont {Davis}}]{Lee2015}%
  \BibitemOpen
  \bibfield  {author} {\bibinfo {author} {\bibfnamefont {I.}~\bibnamefont
  {Lee}}, \bibinfo {author} {\bibfnamefont {C.~K.}\ \bibnamefont {Kim}},
  \bibinfo {author} {\bibfnamefont {J.}~\bibnamefont {Lee}}, \bibinfo {author}
  {\bibfnamefont {S.~J.~L.}\ \bibnamefont {Billinge}}, \bibinfo {author}
  {\bibfnamefont {R.}~\bibnamefont {Zhong}}, \bibinfo {author} {\bibfnamefont
  {J.~A.}\ \bibnamefont {Schneeloch}}, \bibinfo {author} {\bibfnamefont
  {T.}~\bibnamefont {Liu}}, \bibinfo {author} {\bibfnamefont {T.}~\bibnamefont
  {Valla}}, \bibinfo {author} {\bibfnamefont {J.~M.}\ \bibnamefont
  {Tranquada}}, \bibinfo {author} {\bibfnamefont {G.}~\bibnamefont {Gu}},\ and\
  \bibinfo {author} {\bibfnamefont {J.~C.~S.}\ \bibnamefont {Davis}},\ }\href
  {https://doi.org/10.1073/pnas.1424322112} {\bibfield  {journal} {\bibinfo
  {journal} {Proc. Natl. Acad. Sci. U.S.A.}\ }\textbf {\bibinfo {volume}
  {112}},\ \bibinfo {pages} {1316} (\bibinfo {year} {2015})}\BibitemShut
  {NoStop}%
\bibitem [{\citenamefont {Chong}\ \emph {et~al.}(2020)\citenamefont {Chong},
  \citenamefont {Liu}, \citenamefont {Sharma}, \citenamefont {Kostin},
  \citenamefont {Gu}, \citenamefont {Fujita}, \citenamefont {Davis},\ and\
  \citenamefont {Sprau}}]{Chong2020}%
  \BibitemOpen
  \bibfield  {author} {\bibinfo {author} {\bibfnamefont {Y.~X.}\ \bibnamefont
  {Chong}}, \bibinfo {author} {\bibfnamefont {X.}~\bibnamefont {Liu}}, \bibinfo
  {author} {\bibfnamefont {R.}~\bibnamefont {Sharma}}, \bibinfo {author}
  {\bibfnamefont {A.}~\bibnamefont {Kostin}}, \bibinfo {author} {\bibfnamefont
  {G.}~\bibnamefont {Gu}}, \bibinfo {author} {\bibfnamefont {K.}~\bibnamefont
  {Fujita}}, \bibinfo {author} {\bibfnamefont {J.~C.~S.}\ \bibnamefont
  {Davis}},\ and\ \bibinfo {author} {\bibfnamefont {P.~O.}\ \bibnamefont
  {Sprau}},\ }\href@noop {} {\bibfield  {journal} {\bibinfo  {journal} {Nano
  Lett.}\ }\textbf {\bibinfo {volume} {20}},\ \bibinfo {pages} {8001} (\bibinfo
  {year} {2020})}\BibitemShut {NoStop}%
\bibitem [{\citenamefont {Borgwardt}\ \emph {et~al.}(2016)\citenamefont
  {Borgwardt}, \citenamefont {Lux}, \citenamefont {Vergara}, \citenamefont
  {Wang}, \citenamefont {Taskin}, \citenamefont {Segawa}, \citenamefont {van
  Loosdrecht}, \citenamefont {Ando}, \citenamefont {Rosch},\ and\ \citenamefont
  {Gr\"uninger}}]{Borgwardt2016}%
  \BibitemOpen
  \bibfield  {author} {\bibinfo {author} {\bibfnamefont {N.}~\bibnamefont
  {Borgwardt}}, \bibinfo {author} {\bibfnamefont {J.}~\bibnamefont {Lux}},
  \bibinfo {author} {\bibfnamefont {I.}~\bibnamefont {Vergara}}, \bibinfo
  {author} {\bibfnamefont {Z.}~\bibnamefont {Wang}}, \bibinfo {author}
  {\bibfnamefont {A.~A.}\ \bibnamefont {Taskin}}, \bibinfo {author}
  {\bibfnamefont {K.}~\bibnamefont {Segawa}}, \bibinfo {author} {\bibfnamefont
  {P.~H.~M.}\ \bibnamefont {van Loosdrecht}}, \bibinfo {author} {\bibfnamefont
  {Y.}~\bibnamefont {Ando}}, \bibinfo {author} {\bibfnamefont {A.}~\bibnamefont
  {Rosch}},\ and\ \bibinfo {author} {\bibfnamefont {M.}~\bibnamefont
  {Gr\"uninger}},\ }\href {https://doi.org/10.1103/PhysRevB.93.245149}
  {\bibfield  {journal} {\bibinfo  {journal} {Phys. Rev. B}\ }\textbf {\bibinfo
  {volume} {93}},\ \bibinfo {pages} {245149} (\bibinfo {year}
  {2016})}\BibitemShut {NoStop}%
\bibitem [{\citenamefont {Knispel}\ \emph {et~al.}(2017)\citenamefont
  {Knispel}, \citenamefont {Jolie}, \citenamefont {Borgwardt}, \citenamefont
  {Lux}, \citenamefont {Wang}, \citenamefont {Ando}, \citenamefont {Rosch},
  \citenamefont {Michely},\ and\ \citenamefont {Gr\"uninger}}]{Knispel2017}%
  \BibitemOpen
  \bibfield  {author} {\bibinfo {author} {\bibfnamefont {T.}~\bibnamefont
  {Knispel}}, \bibinfo {author} {\bibfnamefont {W.}~\bibnamefont {Jolie}},
  \bibinfo {author} {\bibfnamefont {N.}~\bibnamefont {Borgwardt}}, \bibinfo
  {author} {\bibfnamefont {J.}~\bibnamefont {Lux}}, \bibinfo {author}
  {\bibfnamefont {Z.}~\bibnamefont {Wang}}, \bibinfo {author} {\bibfnamefont
  {Y.}~\bibnamefont {Ando}}, \bibinfo {author} {\bibfnamefont {A.}~\bibnamefont
  {Rosch}}, \bibinfo {author} {\bibfnamefont {T.}~\bibnamefont {Michely}},\
  and\ \bibinfo {author} {\bibfnamefont {M.}~\bibnamefont {Gr\"uninger}},\
  }\href {https://doi.org/10.1103/PhysRevB.96.195135} {\bibfield  {journal}
  {\bibinfo  {journal} {Phys. Rev. B}\ }\textbf {\bibinfo {volume} {96}},\
  \bibinfo {pages} {195135} (\bibinfo {year} {2017})}\BibitemShut {NoStop}%
\bibitem [{\citenamefont {Breunig}\ \emph {et~al.}(2017)\citenamefont
  {Breunig}, \citenamefont {Wang}, \citenamefont {Taskin}, \citenamefont {Lux},
  \citenamefont {Rosch},\ and\ \citenamefont {Ando}}]{Breunig2017}%
  \BibitemOpen
  \bibfield  {author} {\bibinfo {author} {\bibfnamefont {O.}~\bibnamefont
  {Breunig}}, \bibinfo {author} {\bibfnamefont {Z.}~\bibnamefont {Wang}},
  \bibinfo {author} {\bibfnamefont {A.}~\bibnamefont {Taskin}}, \bibinfo
  {author} {\bibfnamefont {J.}~\bibnamefont {Lux}}, \bibinfo {author}
  {\bibfnamefont {A.}~\bibnamefont {Rosch}},\ and\ \bibinfo {author}
  {\bibfnamefont {Y.}~\bibnamefont {Ando}},\ }\href
  {https://doi.org/10.1038/ncomms15545} {\bibfield  {journal} {\bibinfo
  {journal} {Nat. Commun.}\ }\textbf {\bibinfo {volume} {8}},\ \bibinfo {pages}
  {15545} (\bibinfo {year} {2017})}\BibitemShut {NoStop}%
\bibitem [{\citenamefont {Skinner}\ \emph {et~al.}(2013)\citenamefont
  {Skinner}, \citenamefont {T.~Chen},\ and\ \citenamefont
  {Shklovskii}}]{Skinner2013A}%
  \BibitemOpen
  \bibfield  {author} {\bibinfo {author} {\bibfnamefont {B.}~\bibnamefont
  {Skinner}}, \bibinfo {author} {\bibfnamefont {T.}~\bibnamefont {T.~Chen}},\
  and\ \bibinfo {author} {\bibfnamefont {B.~I.}\ \bibnamefont {Shklovskii}},\
  }\href@noop {} {\bibfield  {journal} {\bibinfo  {journal} {J. Exp. Theor.
  Phys.}\ }\textbf {\bibinfo {volume} {117}},\ \bibinfo {pages} {579} (\bibinfo
  {year} {2013})}\BibitemShut {NoStop}%
\bibitem [{\citenamefont {Skinner}\ \emph {et~al.}(2012)\citenamefont
  {Skinner}, \citenamefont {Chen},\ and\ \citenamefont
  {Shklovskii}}]{Skinner2012}%
  \BibitemOpen
  \bibfield  {author} {\bibinfo {author} {\bibfnamefont {B.}~\bibnamefont
  {Skinner}}, \bibinfo {author} {\bibfnamefont {T.}~\bibnamefont {Chen}},\ and\
  \bibinfo {author} {\bibfnamefont {B.~I.}\ \bibnamefont {Shklovskii}},\ }\href
  {https://doi.org/10.1103/PhysRevLett.109.176801} {\bibfield  {journal}
  {\bibinfo  {journal} {Phys. Rev. Lett.}\ }\textbf {\bibinfo {volume} {109}},\
  \bibinfo {pages} {176801} (\bibinfo {year} {2012})}\BibitemShut {NoStop}%
\bibitem [{\citenamefont {Chen}\ and\ \citenamefont
  {Shklovskii}(2013)}]{Chen2013}%
  \BibitemOpen
  \bibfield  {author} {\bibinfo {author} {\bibfnamefont {T.}~\bibnamefont
  {Chen}}\ and\ \bibinfo {author} {\bibfnamefont {B.~I.}\ \bibnamefont
  {Shklovskii}},\ }\href {https://doi.org/10.1103/PhysRevB.87.165119}
  {\bibfield  {journal} {\bibinfo  {journal} {Phys. Rev. B}\ }\textbf {\bibinfo
  {volume} {87}},\ \bibinfo {pages} {165119} (\bibinfo {year}
  {2013})}\BibitemShut {NoStop}%
\bibitem [{\citenamefont {Skinner}\ and\ \citenamefont
  {Shklovskii}(2013)}]{Skinner2013B}%
  \BibitemOpen
  \bibfield  {author} {\bibinfo {author} {\bibfnamefont {B.}~\bibnamefont
  {Skinner}}\ and\ \bibinfo {author} {\bibfnamefont {B.~I.}\ \bibnamefont
  {Shklovskii}},\ }\href {https://doi.org/10.1103/PhysRevB.87.075454}
  {\bibfield  {journal} {\bibinfo  {journal} {Phys. Rev. B}\ }\textbf {\bibinfo
  {volume} {87}},\ \bibinfo {pages} {075454} (\bibinfo {year}
  {2013})}\BibitemShut {NoStop}%
\bibitem [{\citenamefont {B\"omerich}\ \emph {et~al.}(2017)\citenamefont
  {B\"omerich}, \citenamefont {Lux}, \citenamefont {Feng},\ and\ \citenamefont
  {Rosch}}]{Bomerich2017}%
  \BibitemOpen
  \bibfield  {author} {\bibinfo {author} {\bibfnamefont {T.}~\bibnamefont
  {B\"omerich}}, \bibinfo {author} {\bibfnamefont {J.}~\bibnamefont {Lux}},
  \bibinfo {author} {\bibfnamefont {Q.~T.}\ \bibnamefont {Feng}},\ and\
  \bibinfo {author} {\bibfnamefont {A.}~\bibnamefont {Rosch}},\ }\href
  {https://doi.org/10.1103/PhysRevB.96.075204} {\bibfield  {journal} {\bibinfo
  {journal} {Phys. Rev. B}\ }\textbf {\bibinfo {volume} {96}},\ \bibinfo
  {pages} {075204} (\bibinfo {year} {2017})}\BibitemShut {NoStop}%
\bibitem [{\citenamefont {Huang}\ and\ \citenamefont
  {Shklovskii}(2021)}]{Huang2021}%
  \BibitemOpen
  \bibfield  {author} {\bibinfo {author} {\bibfnamefont {Y.}~\bibnamefont
  {Huang}}\ and\ \bibinfo {author} {\bibfnamefont {B.~I.}\ \bibnamefont
  {Shklovskii}},\ }\href {https://doi.org/10.1103/PhysRevB.103.165409}
  {\bibfield  {journal} {\bibinfo  {journal} {Phys. Rev. B}\ }\textbf {\bibinfo
  {volume} {103}},\ \bibinfo {pages} {165409} (\bibinfo {year}
  {2021})}\BibitemShut {NoStop}%
\bibitem [{\citenamefont {Zhang}\ \emph {et~al.}(2011)\citenamefont {Zhang},
  \citenamefont {Chang}, \citenamefont {Zhang}, \citenamefont {Wen},
  \citenamefont {Feng}, \citenamefont {Li}, \citenamefont {Liu}, \citenamefont
  {He}, \citenamefont {Wang}, \citenamefont {Chen}, \citenamefont {Xue},
  \citenamefont {Ma},\ and\ \citenamefont {Wang}}]{Zhang2011}%
  \BibitemOpen
  \bibfield  {author} {\bibinfo {author} {\bibfnamefont {J.}~\bibnamefont
  {Zhang}}, \bibinfo {author} {\bibfnamefont {C.-Z.}\ \bibnamefont {Chang}},
  \bibinfo {author} {\bibfnamefont {Z.}~\bibnamefont {Zhang}}, \bibinfo
  {author} {\bibfnamefont {J.}~\bibnamefont {Wen}}, \bibinfo {author}
  {\bibfnamefont {X.}~\bibnamefont {Feng}}, \bibinfo {author} {\bibfnamefont
  {K.}~\bibnamefont {Li}}, \bibinfo {author} {\bibfnamefont {M.}~\bibnamefont
  {Liu}}, \bibinfo {author} {\bibfnamefont {K.}~\bibnamefont {He}}, \bibinfo
  {author} {\bibfnamefont {L.}~\bibnamefont {Wang}}, \bibinfo {author}
  {\bibfnamefont {X.}~\bibnamefont {Chen}}, \bibinfo {author} {\bibfnamefont
  {Q.-K.}\ \bibnamefont {Xue}}, \bibinfo {author} {\bibfnamefont
  {X.}~\bibnamefont {Ma}},\ and\ \bibinfo {author} {\bibfnamefont
  {Y.}~\bibnamefont {Wang}},\ }\href {https://doi.org/10.1038/ncomms1588}
  {\bibfield  {journal} {\bibinfo  {journal} {Nat. Commun.}\ }\textbf {\bibinfo
  {volume} {2}},\ \bibinfo {pages} {574} (\bibinfo {year} {2011})}\BibitemShut
  {NoStop}%
\bibitem [{\citenamefont {Nandi}\ \emph {et~al.}(2018)\citenamefont {Nandi},
  \citenamefont {Skinner}, \citenamefont {Lee}, \citenamefont {Huang},
  \citenamefont {Shain}, \citenamefont {Chang}, \citenamefont {Ou},
  \citenamefont {Lee}, \citenamefont {Ward}, \citenamefont {Moodera},
  \citenamefont {Kim}, \citenamefont {Halperin},\ and\ \citenamefont
  {Yacoby}}]{Nandi2018}%
  \BibitemOpen
  \bibfield  {author} {\bibinfo {author} {\bibfnamefont {D.}~\bibnamefont
  {Nandi}}, \bibinfo {author} {\bibfnamefont {B.}~\bibnamefont {Skinner}},
  \bibinfo {author} {\bibfnamefont {G.~H.}\ \bibnamefont {Lee}}, \bibinfo
  {author} {\bibfnamefont {K.-F.}\ \bibnamefont {Huang}}, \bibinfo {author}
  {\bibfnamefont {K.}~\bibnamefont {Shain}}, \bibinfo {author} {\bibfnamefont
  {C.-Z.}\ \bibnamefont {Chang}}, \bibinfo {author} {\bibfnamefont
  {Y.}~\bibnamefont {Ou}}, \bibinfo {author} {\bibfnamefont {S.-P.}\
  \bibnamefont {Lee}}, \bibinfo {author} {\bibfnamefont {J.}~\bibnamefont
  {Ward}}, \bibinfo {author} {\bibfnamefont {J.~S.}\ \bibnamefont {Moodera}},
  \bibinfo {author} {\bibfnamefont {P.}~\bibnamefont {Kim}}, \bibinfo {author}
  {\bibfnamefont {B.~I.}\ \bibnamefont {Halperin}},\ and\ \bibinfo {author}
  {\bibfnamefont {A.}~\bibnamefont {Yacoby}},\ }\href
  {https://doi.org/10.1103/PhysRevB.98.214203} {\bibfield  {journal} {\bibinfo
  {journal} {Phys. Rev. B}\ }\textbf {\bibinfo {volume} {98}},\ \bibinfo
  {pages} {214203} (\bibinfo {year} {2018})}\BibitemShut {NoStop}%
\bibitem [{\citenamefont {Tsemekhman}\ \emph {et~al.}(1997)\citenamefont
  {Tsemekhman}, \citenamefont {Tsemekhman}, \citenamefont {Wexler},
  \citenamefont {Han},\ and\ \citenamefont {Thouless}}]{Tsemekhman1997}%
  \BibitemOpen
  \bibfield  {author} {\bibinfo {author} {\bibfnamefont {V.}~\bibnamefont
  {Tsemekhman}}, \bibinfo {author} {\bibfnamefont {K.}~\bibnamefont
  {Tsemekhman}}, \bibinfo {author} {\bibfnamefont {C.}~\bibnamefont {Wexler}},
  \bibinfo {author} {\bibfnamefont {J.~H.}\ \bibnamefont {Han}},\ and\ \bibinfo
  {author} {\bibfnamefont {D.~J.}\ \bibnamefont {Thouless}},\ }\href
  {https://doi.org/10.1103/PhysRevB.55.R10201} {\bibfield  {journal} {\bibinfo
  {journal} {Phys. Rev. B}\ }\textbf {\bibinfo {volume} {55}},\ \bibinfo
  {pages} {R10201} (\bibinfo {year} {1997})}\BibitemShut {NoStop}%
\bibitem [{\citenamefont {Komiyama}\ and\ \citenamefont
  {Kawaguchi}(2000)}]{Komiyama2000}%
  \BibitemOpen
  \bibfield  {author} {\bibinfo {author} {\bibfnamefont {S.}~\bibnamefont
  {Komiyama}}\ and\ \bibinfo {author} {\bibfnamefont {Y.}~\bibnamefont
  {Kawaguchi}},\ }\href {https://doi.org/10.1103/PhysRevB.61.2014} {\bibfield
  {journal} {\bibinfo  {journal} {Phys. Rev. B}\ }\textbf {\bibinfo {volume}
  {61}},\ \bibinfo {pages} {2014} (\bibinfo {year} {2000})}\BibitemShut
  {NoStop}%
\end{thebibliography}

\begin{thebibliography}{12}%
\makeatletter
\providecommand \@ifxundefined [1]{%
 \@ifx{#1\undefined}
}%
\providecommand \@ifnum [1]{%
 \ifnum #1\expandafter \@firstoftwo
 \else \expandafter \@secondoftwo
 \fi
}%
\providecommand \@ifx [1]{%
 \ifx #1\expandafter \@firstoftwo
 \else \expandafter \@secondoftwo
 \fi
}%
\providecommand \natexlab [1]{#1}%
\providecommand \enquote  [1]{``#1''}%
\providecommand \bibnamefont  [1]{#1}%
\providecommand \bibfnamefont [1]{#1}%
\providecommand \citenamefont [1]{#1}%
\providecommand \href@noop [0]{\@secondoftwo}%
\providecommand \href [0]{\begingroup \@sanitize@url \@href}%
\providecommand \@href[1]{\@@startlink{#1}\@@href}%
\providecommand \@@href[1]{\endgroup#1\@@endlink}%
\providecommand \@sanitize@url [0]{\catcode `\\12\catcode `\$12\catcode
  `\&12\catcode `\#12\catcode `\^12\catcode `\_12\catcode `\%12\relax}%
\providecommand \@@startlink[1]{}%
\providecommand \@@endlink[0]{}%
\providecommand \url  [0]{\begingroup\@sanitize@url \@url }%
\providecommand \@url [1]{\endgroup\@href {#1}{\urlprefix }}%
\providecommand \urlprefix  [0]{URL }%
\providecommand \Eprint [0]{\href }%
\providecommand \doibase [0]{https://doi.org/}%
\providecommand \selectlanguage [0]{\@gobble}%
\providecommand \bibinfo  [0]{\@secondoftwo}%
\providecommand \bibfield  [0]{\@secondoftwo}%
\providecommand \translation [1]{[#1]}%
\providecommand \BibitemOpen [0]{}%
\providecommand \bibitemStop [0]{}%
\providecommand \bibitemNoStop [0]{.\EOS\space}%
\providecommand \EOS [0]{\spacefactor3000\relax}%
\providecommand \BibitemShut  [1]{\csname bibitem#1\endcsname}%
\let\auto@bib@innerbib\@empty
\bibitem [{\citenamefont {Chang}\ \emph {et~al.}(2015)\citenamefont {Chang},
  \citenamefont {Zhao}, \citenamefont {Kim}, \citenamefont {Wei}, \citenamefont
  {Jain}, \citenamefont {Liu}, \citenamefont {Chan},\ and\ \citenamefont
  {Moodera}}]{Chang2015Bs}%
  \BibitemOpen
  \bibfield  {author} {\bibinfo {author} {\bibfnamefont {C.-Z.}\ \bibnamefont
  {Chang}}, \bibinfo {author} {\bibfnamefont {W.}~\bibnamefont {Zhao}},
  \bibinfo {author} {\bibfnamefont {D.~Y.}\ \bibnamefont {Kim}}, \bibinfo
  {author} {\bibfnamefont {P.}~\bibnamefont {Wei}}, \bibinfo {author}
  {\bibfnamefont {J.~K.}\ \bibnamefont {Jain}}, \bibinfo {author}
  {\bibfnamefont {C.}~\bibnamefont {Liu}}, \bibinfo {author} {\bibfnamefont
  {M.~H.~W.}\ \bibnamefont {Chan}},\ and\ \bibinfo {author} {\bibfnamefont
  {J.~S.}\ \bibnamefont {Moodera}},\ }\href
  {https://doi.org/10.1103/PhysRevLett.115.057206} {\bibfield  {journal}
  {\bibinfo  {journal} {Phys. Rev. Lett.}\ }\textbf {\bibinfo {volume} {115}},\
  \bibinfo {pages} {057206} (\bibinfo {year} {2015})}\BibitemShut {NoStop}%
\bibitem [{\citenamefont {Kou}\ \emph {et~al.}(2014)\citenamefont {Kou} \emph
  {et~al.}}]{Kou2014s}%
  \BibitemOpen
  \bibfield  {author} {\bibinfo {author} {\bibfnamefont {X.}~\bibnamefont
  {Kou}} \emph {et~al.},\ }\href
  {https://doi.org/10.1103/PhysRevLett.113.137201} {\bibfield  {journal}
  {\bibinfo  {journal} {Phys. Rev. Lett.}\ }\textbf {\bibinfo {volume} {113}},\
  \bibinfo {pages} {137201} (\bibinfo {year} {2014})}\BibitemShut {NoStop}%
\bibitem [{\citenamefont {B\"uttiker}(1988)}]{Buttiker1988s}%
  \BibitemOpen
  \bibfield  {author} {\bibinfo {author} {\bibfnamefont {M.}~\bibnamefont
  {B\"uttiker}},\ }\href {https://doi.org/10.1103/PhysRevB.38.9375} {\bibfield
  {journal} {\bibinfo  {journal} {Phys. Rev. B}\ }\textbf {\bibinfo {volume}
  {38}},\ \bibinfo {pages} {9375} (\bibinfo {year} {1988})}\BibitemShut
  {NoStop}%
\bibitem [{\citenamefont {Wang}\ \emph {et~al.}(2013)\citenamefont {Wang},
  \citenamefont {Lian}, \citenamefont {Zhang},\ and\ \citenamefont
  {Zhang}}]{Wang2013s}%
  \BibitemOpen
  \bibfield  {author} {\bibinfo {author} {\bibfnamefont {J.}~\bibnamefont
  {Wang}}, \bibinfo {author} {\bibfnamefont {B.}~\bibnamefont {Lian}}, \bibinfo
  {author} {\bibfnamefont {H.}~\bibnamefont {Zhang}},\ and\ \bibinfo {author}
  {\bibfnamefont {S.-C.}\ \bibnamefont {Zhang}},\ }\href
  {https://doi.org/10.1103/PhysRevLett.111.086803} {\bibfield  {journal}
  {\bibinfo  {journal} {Phys. Rev. Lett.}\ }\textbf {\bibinfo {volume} {111}},\
  \bibinfo {pages} {086803} (\bibinfo {year} {2013})}\BibitemShut {NoStop}%
\bibitem [{\citenamefont {Nandi}\ \emph {et~al.}(2018)\citenamefont {Nandi},
  \citenamefont {Skinner}, \citenamefont {Lee}, \citenamefont {Huang},
  \citenamefont {Shain}, \citenamefont {Chang}, \citenamefont {Ou},
  \citenamefont {Lee}, \citenamefont {Ward}, \citenamefont {Moodera},
  \citenamefont {Kim}, \citenamefont {Halperin},\ and\ \citenamefont
  {Yacoby}}]{Nandi2018s}%
  \BibitemOpen
  \bibfield  {author} {\bibinfo {author} {\bibfnamefont {D.}~\bibnamefont
  {Nandi}}, \bibinfo {author} {\bibfnamefont {B.}~\bibnamefont {Skinner}},
  \bibinfo {author} {\bibfnamefont {G.~H.}\ \bibnamefont {Lee}}, \bibinfo
  {author} {\bibfnamefont {K.-F.}\ \bibnamefont {Huang}}, \bibinfo {author}
  {\bibfnamefont {K.}~\bibnamefont {Shain}}, \bibinfo {author} {\bibfnamefont
  {C.-Z.}\ \bibnamefont {Chang}}, \bibinfo {author} {\bibfnamefont
  {Y.}~\bibnamefont {Ou}}, \bibinfo {author} {\bibfnamefont {S.-P.}\
  \bibnamefont {Lee}}, \bibinfo {author} {\bibfnamefont {J.}~\bibnamefont
  {Ward}}, \bibinfo {author} {\bibfnamefont {J.~S.}\ \bibnamefont {Moodera}},
  \bibinfo {author} {\bibfnamefont {P.}~\bibnamefont {Kim}}, \bibinfo {author}
  {\bibfnamefont {B.~I.}\ \bibnamefont {Halperin}},\ and\ \bibinfo {author}
  {\bibfnamefont {A.}~\bibnamefont {Yacoby}},\ }\href
  {https://doi.org/10.1103/PhysRevB.98.214203} {\bibfield  {journal} {\bibinfo
  {journal} {Phys. Rev. B}\ }\textbf {\bibinfo {volume} {98}},\ \bibinfo
  {pages} {214203} (\bibinfo {year} {2018})}\BibitemShut {NoStop}%
\bibitem [{\citenamefont {Huang}\ and\ \citenamefont
  {Shklovskii}(2021)}]{Huang2021s}%
  \BibitemOpen
  \bibfield  {author} {\bibinfo {author} {\bibfnamefont {Y.}~\bibnamefont
  {Huang}}\ and\ \bibinfo {author} {\bibfnamefont {B.~I.}\ \bibnamefont
  {Shklovskii}},\ }\href {https://doi.org/10.1103/PhysRevB.103.165409}
  {\bibfield  {journal} {\bibinfo  {journal} {Phys. Rev. B}\ }\textbf {\bibinfo
  {volume} {103}},\ \bibinfo {pages} {165409} (\bibinfo {year}
  {2021})}\BibitemShut {NoStop}%
\bibitem [{\citenamefont {Lee}\ \emph {et~al.}(2015)\citenamefont {Lee},
  \citenamefont {Kim}, \citenamefont {Lee}, \citenamefont {Billinge},
  \citenamefont {Zhong}, \citenamefont {Schneeloch}, \citenamefont {Liu},
  \citenamefont {Valla}, \citenamefont {Tranquada}, \citenamefont {Gu},\ and\
  \citenamefont {Davis}}]{Lee2015s}%
  \BibitemOpen
  \bibfield  {author} {\bibinfo {author} {\bibfnamefont {I.}~\bibnamefont
  {Lee}}, \bibinfo {author} {\bibfnamefont {C.~K.}\ \bibnamefont {Kim}},
  \bibinfo {author} {\bibfnamefont {J.}~\bibnamefont {Lee}}, \bibinfo {author}
  {\bibfnamefont {S.~J.~L.}\ \bibnamefont {Billinge}}, \bibinfo {author}
  {\bibfnamefont {R.}~\bibnamefont {Zhong}}, \bibinfo {author} {\bibfnamefont
  {J.~A.}\ \bibnamefont {Schneeloch}}, \bibinfo {author} {\bibfnamefont
  {T.}~\bibnamefont {Liu}}, \bibinfo {author} {\bibfnamefont {T.}~\bibnamefont
  {Valla}}, \bibinfo {author} {\bibfnamefont {J.~M.}\ \bibnamefont
  {Tranquada}}, \bibinfo {author} {\bibfnamefont {G.}~\bibnamefont {Gu}},\ and\
  \bibinfo {author} {\bibfnamefont {J.~C.~S.}\ \bibnamefont {Davis}},\ }\href
  {https://doi.org/10.1073/pnas.1424322112} {\bibfield  {journal} {\bibinfo
  {journal} {Proc. Natl. Acad. Sci. U.S.A.}\ }\textbf {\bibinfo {volume}
  {112}},\ \bibinfo {pages} {1316} (\bibinfo {year} {2015})}\BibitemShut
  {NoStop}%
\bibitem [{\citenamefont {Chong}\ \emph {et~al.}(2020)\citenamefont {Chong},
  \citenamefont {Liu}, \citenamefont {Sharma}, \citenamefont {Kostin},
  \citenamefont {Gu}, \citenamefont {Fujita}, \citenamefont {Davis},\ and\
  \citenamefont {Sprau}}]{Chong2020s}%
  \BibitemOpen
  \bibfield  {author} {\bibinfo {author} {\bibfnamefont {Y.~X.}\ \bibnamefont
  {Chong}}, \bibinfo {author} {\bibfnamefont {X.}~\bibnamefont {Liu}}, \bibinfo
  {author} {\bibfnamefont {R.}~\bibnamefont {Sharma}}, \bibinfo {author}
  {\bibfnamefont {A.}~\bibnamefont {Kostin}}, \bibinfo {author} {\bibfnamefont
  {G.}~\bibnamefont {Gu}}, \bibinfo {author} {\bibfnamefont {K.}~\bibnamefont
  {Fujita}}, \bibinfo {author} {\bibfnamefont {J.~C.~S.}\ \bibnamefont
  {Davis}},\ and\ \bibinfo {author} {\bibfnamefont {P.~O.}\ \bibnamefont
  {Sprau}},\ }\href@noop {} {\bibfield  {journal} {\bibinfo  {journal} {Nano
  Lett.}\ }\textbf {\bibinfo {volume} {20}},\ \bibinfo {pages} {8001} (\bibinfo
  {year} {2020})}\BibitemShut {NoStop}%
\bibitem [{\citenamefont {Fox}\ \emph {et~al.}(2018)\citenamefont {Fox},
  \citenamefont {Rosen}, \citenamefont {Yang}, \citenamefont {Jones},
  \citenamefont {Elmquist}, \citenamefont {Kou}, \citenamefont {Pan},
  \citenamefont {Wang},\ and\ \citenamefont {Goldhaber-Gordon}}]{Fox2018s}%
  \BibitemOpen
  \bibfield  {author} {\bibinfo {author} {\bibfnamefont {E.~J.}\ \bibnamefont
  {Fox}}, \bibinfo {author} {\bibfnamefont {I.~T.}\ \bibnamefont {Rosen}},
  \bibinfo {author} {\bibfnamefont {Y.}~\bibnamefont {Yang}}, \bibinfo {author}
  {\bibfnamefont {G.~R.}\ \bibnamefont {Jones}}, \bibinfo {author}
  {\bibfnamefont {R.~E.}\ \bibnamefont {Elmquist}}, \bibinfo {author}
  {\bibfnamefont {X.}~\bibnamefont {Kou}}, \bibinfo {author} {\bibfnamefont
  {L.}~\bibnamefont {Pan}}, \bibinfo {author} {\bibfnamefont {K.~L.}\
  \bibnamefont {Wang}},\ and\ \bibinfo {author} {\bibfnamefont
  {D.}~\bibnamefont {Goldhaber-Gordon}},\ }\href
  {https://doi.org/10.1103/PhysRevB.98.075145} {\bibfield  {journal} {\bibinfo
  {journal} {Phys. Rev. B}\ }\textbf {\bibinfo {volume} {98}},\ \bibinfo
  {pages} {075145} (\bibinfo {year} {2018})}\BibitemShut {NoStop}%
\bibitem [{\citenamefont {Komiyama}\ and\ \citenamefont
  {Kawaguchi}(2000)}]{Komiyama2000s}%
  \BibitemOpen
  \bibfield  {author} {\bibinfo {author} {\bibfnamefont {S.}~\bibnamefont
  {Komiyama}}\ and\ \bibinfo {author} {\bibfnamefont {Y.}~\bibnamefont
  {Kawaguchi}},\ }\href {https://doi.org/10.1103/PhysRevB.61.2014} {\bibfield
  {journal} {\bibinfo  {journal} {Phys. Rev. B}\ }\textbf {\bibinfo {volume}
  {61}},\ \bibinfo {pages} {2014} (\bibinfo {year} {2000})}\BibitemShut
  {NoStop}%
\bibitem [{\citenamefont {Nachtwei}(1999)}]{Nachtwei1999s}%
  \BibitemOpen
  \bibfield  {author} {\bibinfo {author} {\bibfnamefont {G.}~\bibnamefont
  {Nachtwei}},\ }\href {https://doi.org/10.1016/S1386-9477(98)00251-3}
  {\bibfield  {journal} {\bibinfo  {journal} {Physica E Low Dimens. Syst.
  Nanostruct.}\ }\textbf {\bibinfo {volume} {4}},\ \bibinfo {pages} {79}
  (\bibinfo {year} {1999})}\BibitemShut {NoStop}%
\bibitem [{\citenamefont {Jeckelmann}\ and\ \citenamefont
  {Jeanneret}(2001)}]{Jeckelmann2001s}%
  \BibitemOpen
  \bibfield  {author} {\bibinfo {author} {\bibfnamefont {B.}~\bibnamefont
  {Jeckelmann}}\ and\ \bibinfo {author} {\bibfnamefont {B.}~\bibnamefont
  {Jeanneret}},\ }\href {https://doi.org/10.1088/0034-4885/64/12/201}
  {\bibfield  {journal} {\bibinfo  {journal} {Rep. Prog. Phys.}\ }\textbf
  {\bibinfo {volume} {64}},\ \bibinfo {pages} {1603} (\bibinfo {year}
  {2001})}\BibitemShut {NoStop}%
\end{thebibliography}
\end{document}